\documentclass[showpacs,twocolumn,amsmath,amssymb,prb,superscriptaddress]{revtex4}

\usepackage{graphicx}
\usepackage{amsmath,amsfonts,amssymb,amsthm}
\usepackage{fancyhdr}
\usepackage{mathrsfs}
\usepackage{epstopdf}
\usepackage{setspace}
\usepackage{hyperref}

\begin{document}



\title{Correlated Cooper pair transport and microwave photon emission in \\
the dynamical Coulomb blockade}

\author{Juha Lepp\"akangas}

\affiliation{Microtechnology and Nanoscience, MC2, Chalmers
University of Technology, SE-412 96 G\"oteborg, Sweden}

\affiliation{Institut f\"ur Theoretische Festk\"orperphysik, Karlsruhe Institute of Technology, D-76128 Karlsruhe, Germany}

\author{Mikael Fogelstr\"om}

\affiliation{Microtechnology and Nanoscience, MC2, Chalmers
University of Technology, SE-412 96 G\"oteborg, Sweden}

\author{Michael Marthaler}

\affiliation{Institut f\"ur Theoretische Festk\"orperphysik, Karlsruhe Institute of Technology, D-76128 Karlsruhe, Germany}

\author{G\"oran Johansson}

\affiliation{Microtechnology and Nanoscience, MC2, Chalmers
University of Technology, SE-412 96 G\"oteborg, Sweden}



\begin{abstract}
We study theoretically electromagnetic radiation emitted by inelastic Cooper-pair tunneling. 
We consider a dc-voltage-biased superconducting transmission line terminated by a Josephson junction.
We show that the generated continuous-mode electromagnetic field can be expressed as a function of the time-dependent current across the Josephson junction.
The leading-order expansion in the tunneling coupling, similar to the "$P(E)$-theory",
has previously been used to investigate the photon emission statistics in the limit of sequential (independent) Cooper-pair tunneling.
By explicitly evaluating the system characteristics up to the fourth-order in the tunneling coupling,
we account for dynamics between consecutively tunneling Cooper pairs.
Within this approach we investigate how temporal correlations in the charge transport can be seen in the first- and second-order coherences of the emitted microwave radiation.
\end{abstract}

\maketitle

\section{Introduction}
Charge transport across a mesoscopic constriction is affected by the electromagnetic properties of the biasing circuitry.
For example, charge tunneling between two conductors 
can be enhanced or inhibited, depending on which electromagnetic modes the tunneling charge is able to excite~\cite{Devoret1990,Girvin1990,Ingold1992,Holst1994,Hofheinz2011}.
If the tunnel junction is superconducting, the energy emitted to the electromagnetic environment can be controlled by an external bias,
due to the absence of electronic states below the superconducting energy gap.
The possibility to engineer the electromagnetic environment and to guide the emitted microwaves as wanted,
lends this system to a versatile source for propagating 
microwave radiation~\cite{Hofheinz2011,Leppakangas2013,Ulm2013,Nottingham2013,Leppakangas20152,Ulm2015,Paris2015,Dambach2015}.
Also motivated by fundamental understanding of quantum transport, the quantum optical aspects of mesoscopic charge
conduction are presently under active 
theoretical~\cite{Beenakker2001,Beenakker2004,Marthaler2011,Jin2011,Nazarov2012,Leppakangas2013,Leppakangas20152,Ulm2013,Nottingham2013,Leppakangas2014,NatureCommunications2014,Jin2014,Portier2,Ulm2015,Paris2015,Cottet2015,Dambach2015,Mendes2015,Quassemi2015,Hassler2015} and experimental~\cite{Bajjani2010,Pashkin2011,Hofheinz2011,Delbecq2013,Reulet1,Reulet2,Reulet3,Portier2014,Rimberg2014,Petta2015} research.

Photon-assisted charge-transport across low-transparency tunnel junctions
has been well understood within the framework of the $P(E)$-theory~\cite{Devoret1990,Girvin1990,Ingold1992,Holst1994}.
It describes effects such as multi-photon-assisted charge tunneling and the Coulomb blockade.
Recently, it has been understood that the theory also characterizes  the simultaneously emitted electromagnetic radiation~\cite{Hofheinz2011,Leppakangas2013},
analogously predicting generation of non-classical photon pairs~\cite{Leppakangas2013} and anti-bunched photons~\cite{Leppakangas20152}. 
These non-classical states of light are important tools for quantum-information applications.
The versatile theory captures the reaction of an arbitrary continuous-mode environment
to a single charge-tunneling event exactly.
However, this perturbative approach is equivalent to the Fermi's golden rule and cannot model dynamics beyond the independent tunneling limit.
In specific situations we can straightforwardly study effects beyond this  by using Jaynes-Cummings type approaches,
such as in the case of a single-mode electromagnetic environment.
Here, it has been predicted that far from equilibrium a simple dc-bias can drive the cavity into rather exotic states, for example,
exhibiting squeezing~\cite{Nottingham2013}, sub-Poissonian photon distributions~\cite{Ulm2013} and multi-stable trapping~\cite{Marthaler2011}.

An important characteristic of creating electromagnetic radiation solely by inelastic charge transport
is the usual rather fast dephasing of the emitted field.
This is due to the effective series resistive environment, which is needed to create high-frequency radiation
but, in the presence of finite zero-frequency resistivity,
will also inevitably induce low-frequency voltage-fluctuations, due to a finite temperature or shot noise
in the charge transport~\cite{Leppakangas2014,Koch1,Koch2,Likharev}.
It is therefore essential to carefully consider the effect of both low- and high-frequency properties of the biasing circuitry to the studied phenomena.
This is automatically addressed by $P(E)$-type approaches studied here.

\begin{figure}[tb!]
\includegraphics[width=0.95\linewidth]{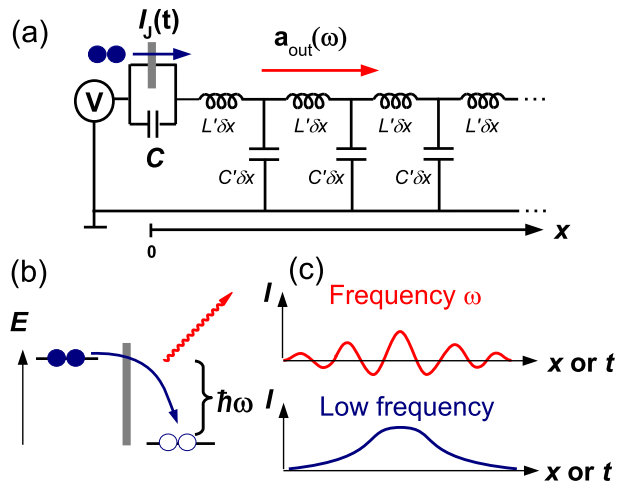}
\caption{
(a) We consider a voltage $V$ biased Josephson junction in parallel with a junction capacitance $C$ and in series with
a transmission line (TL),
that is here represented as an infinite set of capacitances $C'$ and inductances $L'$  per unit length.
(b) Charge tunneling is influenced by the possibility to emit radiation in the TL.
In the superconducting state, 
this is the only way for a tunneling Cooper pair to dissipate the gained electrostatic energy.
(c) In the considered theory circuit, the tunneled charge and the emitted photon(s) can be seen as
excitations in the low- and high-frequency domains of the same electromagnetic environment, respectively.
}
\label{fig:Intro}
\end{figure}

In this article, we establish a continuous-mode description of the electromagnetic radiation emitted by inelastic Cooper-pair tunneling
which accounts for correlations between consecutively tunneling Cooper pairs.
We study relations between the Cooper-pair current and microwave photon emission in
a recently experimentally realized setup~\cite{Hofheinz2011},
where a dc-voltage biased transmission line is terminated by a Josephson junction,  see Fig.~\ref{fig:Intro}.
This is equivalent to a Josephson junction that is voltage-biased in series with a resistive environment~\cite{Ingold1992}.
In this setup the outgoing direct current and high-frequency microwave radiation
can be guided to different routes  and measured independently.
With a proper choice of its parameters the transmission line can, in princible, describe any electromagnetic environment of the Josephson junction,
and is therefore an excellent system for understanding photon-assisted Cooper-pair tunneling.
We find that the formalism allows for investigation of the emitted radiation in terms of the tunneling current across the junction.
This is used to show that the finite-frequency current noise at the junction
is up to a linear filter function and a term describing thermal noise determined by the outgoing photon flux.
The flux-flux correlations of the emitted field are, however,
also functions of other moments of the junction current.
The leading-order expansion in the tunneling coupling, similar to the $P(E)$-theory,
has been previously used to study emission statistics in the independent-tunneling picture~\cite{Leppakangas2013,Leppakangas2014}.
We here demonstrate a systematic expansion of system properties up to the fourth order in the tunneling coupling
and capture temporal correlations between consecutively tunneling Cooper pairs.
A central result is the natural appearance of the Keldysh time-ordering in the photonic correlators,
connecting the normal time-ordering used in quantum optics and the Keldysh ordering in quantum transport.
Within this approach, we are able to study effects in temporal correlations
for all forms of the electromagnetic environment in the limit of weak Cooper-pair tunneling.
Particularly, we study how the correlations in the charge transport affect the first- and
second-order coherences of emitted photons.
The general motivation is to understand how microwave measurements can be used to probe charge-transport statistics, and vice versa,
how correlations in the charge transport can be used to produce non-conventional states of light.

The article is organized as follows. In Section~\ref{sec:Model}, we introduce our theory
for charge transport and the microwave emission in this system.
In Section~\ref{sec:solution}, we derive a description for the emitted field in terms of the junction current
 and study general relations between the finite-frequency noise at the junction and the emitted photon flux density.
 In Section~\ref{sec:power},  they are explicitly calculated up to fourth order in the tunneling coupling across the Josephson junction.
 The calculation is first presented on the Keldysh contour and numerical results for specific transmission lines are analyzed.
 In Section~\ref{sec:Photons}, we do a similar analysis for the second-order coherence
 of the emitted photons. 
Conclusions and outlook are given in Section~\ref{sec:discussion}.

\section{The model}\label{sec:Model}
The system we study consists of a dc-voltage-biased Josephson junction in series with a semi-infinite transmission line (TL), see Fig.~\ref{fig:Intro}.
With a proper choice of the TL parameters it can practically describe any electromagnetic environment of the junction.
Here, for simplicity, we consider an Ohmic TL, which means a constant
capacitance and inductance  per unit length.
The modifications to formulas in a more general case are considered in Appendix~A.
This theory can also straightforwardly be adapted to normal-state junctions~\cite{Quassemi2015}
and therefore goes beyond Josephson junctions.


\subsection{Hamiltonian}
We start from a Hamiltonian that describes the lumped-element circuit in Fig.~\ref{fig:Intro}(a),
\begin{equation}\label{eq:HamiltonianTotal}
H=H_{\rm EE}+ H_{\rm J}.
\end{equation}
The total Hamiltonian $H$ is a sum of the electromagnetic environment Hamiltonian $H_{\rm EE}$ and the Josephson junction Hamiltonian $H_{\rm J}$.
The electromagnetic-environment Hamiltonian has the form (Ohmic TL)
\begin{equation}\label{eq:HamiltonianEnv}
H_{\rm EE}=\frac{\hat Q^2}{2C}+\sum_{n=1}^{\infty}\left[ \frac{\hat q_n^2}{2C'\delta x} +\frac{1}{2L' \delta x}\left(\hat\Phi_{n+1}-\hat\Phi_n\right)^2 \right].
\end{equation}
Here $C$ is the junction capacitance, $C'$ and  $L'$ are the TL capacitance and inductance per unit length,
and $\delta x$ is an infinitesimal segment of the TL.
Each node (island between lumped circuit elements) is associated with an index $n$ and a magnetic flux $\Phi_n$, through which the
electric and magnetic energies per segment $\delta x$ are expressed (the two terms inside the main parentheses, respectively).
The charge and the flux operators at the junction are conjugated variables $[\hat\Phi_1,\hat Q]=2ie$
and similarly for the other nodes $[\hat\Phi_n,\hat q_n]=2ie$.
All other commutators vanish.

The Hamiltonian $H_{\rm J}$ describes the tunnel connection between the
center conductor and ground plane of the superconducting TL (the junction capacitance is accounted for by $H_{\rm EE}$).
Under the assumption $eV<2\Delta$, where $\Delta$ is the energy gap of the superconductor,
we may neglect the quasiparticle degrees of freedom from the analysis. 
Here, the junction Hamiltonian reduces to the usual description of Cooper-pair tunnelling across the Josephson junction,
\begin{equation}\label{eq:HamiltonianJunction}
H_{\rm J}=-E_{\rm J}\cos\left(\omega_{\rm J}t-\hat\phi \right),
\end{equation}
where $E_{\rm J}=(\hbar/2e)I_{\rm c}$ is the Josephson coupling energy (tunneling coupling) and $I_{\rm c}$ the critical current.
The Josephson frequency  $\omega_{\rm J}/2\pi=2eV/h$ accounts for the voltage bias $V$, and
\begin{equation}
\hat \phi= 2\pi\frac{\hat\Phi_1}{\Phi_0},
\end{equation}
is the phase at the Josephson junction and $\Phi_0=h/2e$.

\subsection{Current across the Josephson junction}
In the following, we take use of the time-dependence of the current across the Josephson junction.
This is expressed conveniently in the eigenbasis of $H_{\rm EE}$ treating the junction Hamiltonian $H_{\rm J}$
through a time-ordered expansion. The eigenbasis of $H_{\rm EE}$ corresponds to scattering states of photons for $E_{\rm J}=0$, solved below.
In this approach, the junction current operator at time $t$ has the general form (in the Heisenberg picture)
\begin{equation}\label{eq:MeanCurrent}
\hat I_{\rm J}(t)=I_{\rm c}\sin\left[ \omega_{\rm J}t - \hat \phi(t) \right]=\hat U^{\dagger}(t,t_0)\ \hat I_{\rm J}^0(t) \ \hat U(t,t_0).
\end{equation}
Here $\hat I_{\rm J}^0(t)$ is the current-operator with time evolution given by $H_{\rm EE}$ (we name this free evolution),
\begin{equation}\label{eq:JunctionCurrent0}
\hat I_{\rm J}^0(t)= I_{\rm c}\sin\left[ \omega_{\rm J}t - \hat \phi_0(t) \right].
\end{equation}
The solution for the free evolution of the phase operator $\hat \phi_0(t)$ is given in Eq.~(\ref{PhaseDifference0}).
The time-evolution operator itself has the form,
\begin{equation}\label{eq:TimeEvolution}
\hat U(t,t_0)={\cal T } \exp\left\{ \frac{i}{\hbar}\int_{t_0}^{t} dt' H_{\rm J}(t') \right\}.
\end{equation}
The time evolution of $H_{\rm J}(t')$ is again the free evolution.

The average junction current corresponds to the expectation value $I=\left\langle \hat I_{\rm J}(t)\right\rangle$,
which means tracing out the electromagnetic degrees of freedom.
This has been studied, for example, in Refs.~[\onlinecite{Ingold1992,Ingold1998,Ingold1999}]
assuming thermal equilibrium of the electromagnetic modes at the initial time $t_0\rightarrow -\infty$,
similarly as will also be assumed here.

\subsection{Emitted electromagnetic field}\label{sec:EMField}
The purpose of this article is to investigate properties of the emitted electromagnetic field.
This is conveniently done by considering the Heisenberg equations of motion in the TL.
Taking the continuum limit $\delta x \rightarrow 0$ [see Eq.~(\ref{eq:HamiltonianEnv})], the solution for the magnetic-flux field in the TL 
can be written as~\cite{Wallquist2006,Leppakangas2014}
\begin{eqnarray}\label{eq:WaveGeneral}
&&\hat \Phi(x,t)=\sqrt{\frac{\hbar Z_{0}}{4\pi}}\int_0^\infty\frac{d\omega}{\sqrt{\omega}}\times \\
&\times&\left[ \hat a_{\rm  in} (\omega)e^{-i(k_\omega x+\omega t)}+\hat a_{\rm out} (\omega)e^{-i(-k_\omega x+\omega t)}+{\rm H.c.} \right].  \nonumber
\end{eqnarray}
Here $\hat a_{\rm in}^{(\dagger)}(\omega)$  is the annihilation (creation) operator of the continuous-mode incoming wave
of frequency $\omega$.
It satisfies the standard commutation relation $\left[ \hat a_{\rm in}(\omega),\hat a_{\rm in}^{\dagger}(\omega')  \right]=\delta(\omega-\omega')$.
A similar relation is also valid for the outgoing field.  
The characteristic impedance of the free space is $Z_0=\sqrt{L'/C'}$ and the wave number $k_{\omega}=\omega\sqrt{C'L'}$.

The Heisenberg equation of motion at the junction
takes the form of a boundary condition~\cite{Leppakangas2014,Leppakangas20142},
\begin{eqnarray}\label{eq:BoundaryCondition}
C\ddot{\hat\Phi}(0,t)&-&\frac{1}{L'}\frac{\partial\hat\Phi(x,t)}{\partial x}\vert_{x=0}=\hat I_{\rm J}(t).
\end{eqnarray}
This manifests current conservation at the junction
and connects the annihilation and creation operators of the incoming and outgoing photons.
As the left-hand side of the condition is linear, the difficulty is the treatment of the right-hand side term, the tunneling current.

\section{General relations between the tunneling current and the emitted field}\label{sec:solution}
Here, we derive a general solution for the generated propagating electromagnetic field.
Our tactics is to express the field in terms of the time-dependent current across the Josephson junction,
which in turn can be fully defined through the incoming field.
After this we use this connection to study general relations between junction-current fluctuations 
and the emitted electromagnetic field.

\subsection{Solution}
There are several ways to derive a solution for the out-field as a function of the in-field and their interaction at the boundary.
In literature, this is usually done in general terms by considering a free evolution solution for field operators in the far past (input),
subjected to forward time-evolution with a certain interaction term, and then compared to
freely evolving field operators in the far future (output), see for example Refs.~[\onlinecite{WallsMilburn,QuantumNoiseBook}].
Here, we derive the solution for this specific problem in a more transparent way by explicitly considering leftwards and rightwards propagating photon fields
in the neighbourhood of the tunneling current and the junction capacitor.

The first step towards the solution is to account for that there are no reflections
inside the semi-infinite  TL. The scattering occurs only at the
Josephson junction (TL with resonance structure is analyzed in Appendix A).
This means that the in-field is independent of the out-field.
Furthermore, the in-field is assumed to be in thermal equilibrium and we therefore know its statistics exactly.
The generated out-field can then be treated as an expansion in powers of the (in-field and) tunneling coupling $E_{\rm J}$,
\begin{equation}\label{eq:FormalSolution}
\hat a_{\rm out}(\omega)=\hat a_0(\omega)+\sum_{n\geq 1} \hat a_n(\omega) .
\end{equation}
Here $n$ refers to the power in the tunneling coupling. 
Setting the coupling to zero ($E_{\rm J}=0$)
and Fourier transforming Eq.~(\ref{eq:BoundaryCondition}) one obtains the zeroth-order field operator
\begin{eqnarray}\label{eq:PhaseShift}
\hat a_0(\omega)=\frac{ A(\omega)}{ A^*(\omega)}\hat a_{\rm in}(\omega).
\end{eqnarray}
Here $A(\omega)$ describes the response of the Ohmic TL
\begin{equation}
A(\omega)=\frac{1}{1-i\omega Z_0 C}.
\end{equation}
We see that in the absence of the tunneling current the out-field differs from the in-field by a phase shift
induced by the junction capacitor.
(In comparison to Ref.~[\onlinecite{Leppakangas2014}] we mark $\bar A(\omega)$ here as $A(\omega)$.)

The Cooper-pair tunneling is described by the right-hand-side of the boundary condition, Eq.~(\ref{eq:BoundaryCondition}).
In the out-field, this is accounted for by higher-order terms. Inserting the expansion of Eq.~(\ref{eq:FormalSolution})
to the left-hand side of Eq.~(\ref{eq:BoundaryCondition}), and Fourier transforming, one obtains by solving order by order,
\begin{eqnarray}
a_n(\omega)=i \sqrt{\frac{Z_0}{\pi\hbar\omega}} A(\omega)  \int_{-\infty}^{\infty}e^{i\omega t}dt\left\{ \hat I_{\rm J}(t) \right\}_{n}.
\nonumber
\end{eqnarray}
Here  the formal operation $\{ \cdot \}_n$ picks out $n$:th order term in the tunnelling coupling ($n\geq 1$).
In the next step we take use of the solution for the time evolution at the tunnel junction,
Eq.~(\ref{eq:MeanCurrent}), which provides an explicit expansion in powers of the tunnelling coupling $E_{\rm J}$.
We can now perform a resummation to all orders and obtain
\begin{eqnarray}\label{eq:OutputSolution}
\hat a_{\rm out}(\omega)&=& \hat a_{0}(\omega) + i \sqrt{\frac{Z_0}{\pi\hbar\omega}} A(\omega) \int_{-\infty}^{\infty}dte^{i\omega t}\nonumber\\
&\times& \hat U^{\dagger}(t,-\infty)\  \hat I_{\rm J}^0(t)\ \hat U(t,-\infty).
\end{eqnarray}
The result states that current fluctuations at the constriction linearly define the emitted field.
However, it is important to notice that the structure of the electromagnetic environment
can already essentially modify the junction current fluctuations $I_{\rm J}(t)$.

Finally, the free-evolution of the phase difference at the junction [an operator expanded in Eq.~(\ref{eq:TimeEvolution})] can be expressed via the in-field as~\cite{Leppakangas2014}
\begin{eqnarray}\label{PhaseDifference0}
\hat\phi_0(t) = \frac{ \sqrt{4\pi\hbar Z_0} }{\Phi_0}\int_{0}^{\infty}  \frac{d\omega}{\sqrt{\omega}}  A(\omega)\hat  a_{\rm in}(\omega)e^{-i\omega t}+{\rm H.c.}
\end{eqnarray}
This is a summation over a continuous set of bosonic modes in the TL.

We check the consistency of the solution in Eq.~(\ref{eq:OutputSolution}) with the one
presented in Refs.~[\onlinecite{Leppakangas2013,Leppakangas2014}] (expansion up to second-order in $E_{\rm J}$)
by verifying that all the results derived there are reproduced by the solution in Eq.~(\ref{eq:OutputSolution}).
However, here an explicit time ordering appears, and the solution is free of singularities,
which turn out to be crucial properties when applying the theory beyond the leading order.

\subsection{Connection between emitted field, junction current, and  junction voltage}
So far we have derived that the outgoing field can be expressed as a function of the junction current,
\begin{eqnarray}\label{eq:Connection1}
\hat a_{\rm out}(\omega)=\hat a_{0}(\omega) + i \sqrt{\frac{Z_0}{\pi\hbar\omega}}  A(\omega) \int_{-\infty}^{\infty}dt e^{i\omega t} \hat I_{\rm J}(t).
\end{eqnarray}
This is at the heart of the input-output theory: the time-integrated boundary condition gives the difference between the incoming and outgoing fields~\cite{WallsMilburn,QuantumNoiseBook}.
In this article, when considering the output radiation power, we neglect the zeroth-order contribution $\hat a_{0}(\omega)$, which means here neglecting
(inelastic) redirection of thermal radiation. This has been analyzed in the leading order in Ref.~[\onlinecite{Leppakangas2014}].
This analysis can also be generalized to higher orders, similarly as done in Section~\ref{sec:power}.
We will consider its contribution again in Section~\ref{sec:Photons} where its description of vacuum fluctuations plays an essential role.

We will now investigate the connection between the junction current, junction voltage, and the emitted field in more detail.
The voltage across the Josephson junction operator can be expressed as
\begin{equation}\label{eq:VoltageGeneral}
\hat V_{\rm J}(t)= V-\dot{\hat{\Phi}}(0,t)\equiv V-\delta\hat V_{\rm J}(t).
\end{equation}
To evaluate the above time derivative of the magnetic flux we use~\cite{Leppakangas2014},
\begin{eqnarray}\label{eq:GeneralPhase}
\hat\phi_n(t)= \frac{1}{\Phi_0} \int_{0}^{\infty} d\omega\xi(\omega)\hat a_n(\omega) e^{-i\omega t}+{\rm H.c.}
\end{eqnarray}
Here $\hat\phi_n$ is the $n$:th-order contribution ($n\geq 1$) of the phase at the Josephson junction, $\hat\phi=\hat\phi_0+\sum_{n=1}^{\infty}\hat\phi_n$.
We then obtain a relation between the out-field and time dependence of the junction voltage
\begin{eqnarray}\label{eq:Connection2}
\hat a_{\rm out}(\omega)=\frac{i}{\omega\xi(\omega)} \int_{-\infty}^{\infty}dt e^{i\omega t} \delta\hat V_{\rm J}(t) .
\end{eqnarray}
The Ohmic TL is characterized by $\xi(\omega)=\sqrt{\hbar Z_0\pi/\omega}$.
The different frequency dependence of the front factors on the right-hand sides of Eqs.~(\ref{eq:Connection1}) and~(\ref{eq:Connection2})
reflects the fact that fluctuations in the junction current and junction voltage are filtered differently by the nearby electromagnetic environment.

For a finite bias voltage $V$ there will be a net current $I$ across the junction.
This can also be expanded in powers of the tunnelling coupling~\cite{Ingold1998,Ingold1999},
\begin{equation}
I=\left\langle \hat I_{\rm J}(t) \right\rangle=\sum_{n=1}^{\infty}I_{n}.
\end{equation}
Here we again label each contribution according to its power in $E_{\rm J}$.
Similarly, using equation~(\ref{eq:Connection1}),
we can write for different-order contributions in the out-field,
\begin{eqnarray}
\left\langle \hat a_{\rm out}(\omega)  \right\rangle &=& \sum_{n=0}^{\infty}\left\langle \hat a_n(\omega) \right\rangle = \sum_{n=1}^{\infty}\left\langle \hat a_{n}(\omega) \right\rangle \\
&=& i \sqrt{\frac{4\pi Z_0}{\hbar\omega}} A(\omega) \left( I_{1}+I_{2}+\ldots \right) \delta(\omega). \nonumber
\end{eqnarray}
Only the zero-frequency field contributes since the in-field has no phase coherence. We can write
\begin{equation}
 \left\langle \hat a_{n}(\omega)\right\rangle= i  \sqrt{\frac{4\pi Z_0}{\hbar\omega}} I_{n} \delta(\omega).
\end{equation}
Here $n$ is a positive integer. We insert this into the right-hand side of the time derivative of equation~(\ref{eq:WaveGeneral}), and using Eq.~(\ref{eq:VoltageGeneral}) we get,
\begin{equation}\label{eq:VoltageReduction}
\delta V_{n}=I_{n}Z_0.
\end{equation}
Here $\delta V_{n}$ is the $n$:th order contribution for voltage fluctuations at zero frequency.
Eq.~(\ref{eq:VoltageReduction}) is the Ohm's law for average voltage reduction across the junction due to charge transport in the TL.

\begin{figure}[bt]
\includegraphics[width=0.8\linewidth]{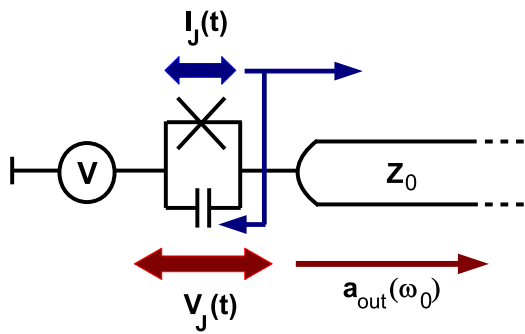}
\caption{
Fluctuations at the Josephson junction always induce radiation to the transmission line ($Z_0$).
The spectrum of the junction-current and junction-voltage fluctuations, and the spectrum of the emitted radiation
are equivalent up to a frequency dependent constant.
However, their relative magnitude might be essentially different, for example,
due to the possibility of fast junction current fluctuations to be shunted by the parallel capacitor.
Also, substancial enhancement of junction current fluctuations
can be induced at certain frequencies by reflections in the transmission line, considered in Section~\ref{sec:Photons}.}
\label{fig:VoltagevsCurrent}
\end{figure}

\subsection{Connection between the spectral densities}\label{sec:SpectrumGeneral}
Current fluctuations at the tunnel junction can be characterized by the correlator
\begin{equation}\label{eq:CurrentCorrelator}
\left\langle \hat I_{\rm J}(t') \hat I_{\rm J}(t) \right\rangle.
\end{equation}
In the considered case this expression will only depend on the difference $\tau=t'-t$.
This is since thermal and quantum low-frequency voltage fluctuations will eventually dephase the junction current.
In this case the following applies
\begin{eqnarray}
&&  \int_{-\infty}^{\infty}dt' e^{-i\omega' t'} \int_{-\infty}^{\infty}dt e^{i\omega t} \left\langle \hat I_{\rm J}(t')\hat I_{\rm J}(t) \right\rangle\nonumber \\
& =&2\pi\delta(\omega-\omega')  \int_{-\infty}^{\infty}d\tau e^{-i\omega\tau}\left\langle \hat I_{\rm J}(\tau)\hat I_{\rm J}(0) \right\rangle\nonumber\\
&\equiv& 2\pi\delta(\omega-\omega') \left\langle \hat I_{\rm J}(\tau)\hat I_{\rm J}(0)\nonumber \right\rangle_{\omega}.
\end{eqnarray}
Using Eq.~(\ref{eq:Connection1})
we relate this to the photon flux density~\cite{Loudon},
\begin{eqnarray}\label{eq:CurrentFluctuations}
f(\omega) &=& \sum_{n,m>0}\int \frac{d\omega'}{2\pi} \left\langle \hat a^{\dagger}_n(\omega) \hat a_m(\omega')   \right\rangle\nonumber\\
&=&\frac{Z_0}{\pi\hbar\omega} \vert A(\omega)\vert^2 \left\langle \hat I_{\rm J}(\tau)\hat I_{\rm J}(0) \right\rangle_{\omega}.
\end{eqnarray}
The spectral densities of the junction current and outgoing radiation are equal
up to the filter $\vert A(\omega)\vert^2$. In analogy to Ref.~[\onlinecite{Ingold1992}],
we can identify the real part of the impedance seen by the tunnel junction,
\begin{equation}\label{eq:Impedance}
{\rm Re}[Z_{\rm t}(\omega)]\equiv Z_0 \vert A(\omega)\vert^2.
\end{equation}
Therefore, photon emission at specific frequencies can be enhanced or inhibited
by the specific design of the electromagnetic environment.

Similarly, using Eq.~(\ref{eq:Connection2}), we obtain for the spectrum of the voltage fluctuations at the junction,
\begin{equation}\label{eq:VoltgeFluctuations}
f(\omega)=\frac{\left\langle \delta\hat V_{\rm J}(\tau)\delta\hat V_{\rm J}(0) \right\rangle_{\omega}}{\omega^2\vert\xi(\omega)\vert^2}.
\end{equation}
Therefore, both the junction current and voltage fluctuation spectrum are proportional to the photon flux, or
the radiation power $s(\omega)=\hbar\omega f(\omega)$.
The difference in the proportionality factors  stems from different filtering
of the corresponding fluctuations by the junction capacitance (and by
possible reflections in the transmission line), as visualized in Fig.~\ref{fig:VoltagevsCurrent}.
These results imply that when studied only through spectral densities,
there is in principle no difference between the statistics of tunneling current and emitted electromagnetic radiation.



\section{Calculation of spectral densities}\label{sec:power}
In this section, we describe how to evaluate power-spectral densities
as perturbation series in the tunneling coupling $E_{\rm J}$.
The calculation is presented on the Keldysh contour.
After this we analyze results in the second and in the fourth order.
The latter is the leading order accounting for dynamics between consecutively tunneling Cooper pairs.
We study the emerging correlation effects also via the recently introduced second-order coherence function for charge transport~\cite{GermanGuys}.


\subsection{Representation on the Keldysh contour}\label{sec:KeldyshCalculation}
The calculation is based on expanding
the time dependence of the Josephson-current operators $\hat I_{\rm J}(t)$ in terms of the time-evolution operators
[see Eqs.~(\ref{eq:MeanCurrent}-\ref{eq:TimeEvolution})],
\begin{eqnarray}\label{eq:CorrelatorDefinition}
&&\left\langle \hat I_{\rm J}(t)\hat I_{\rm J}(t') \right\rangle = \\
&&\left\langle \hat U^{\dagger}(t,-\infty)\ \hat I_{\rm J}^0(t) \ \hat U(t,t') \ \hat I_{\rm J}^0(t') \  \hat U(t',-\infty) \right\rangle  \nonumber.
\end{eqnarray}
We formulate the calculation on the Keldysh contour~\cite{SchoellerSchoen}, which means 
that we represent each term in the expansion as a diagram.
The following set of rules guides how to evaluate the contribution from each topologically different diagram.
We consider the calculation of the power spectrum up to fourth order in the tunneling coupling,
but it is possible to generalize this type of an expansion also to arbitrary orders~\cite{Michael2015}.
Numerical results for Ohmic TLs are presented in Sections~\ref{sec:IndependentTunneling} and \ref{sec:DependentTunneling}.

\subsubsection{Placing the Cooper-pair tunneling events}
\begin{figure}[bt]
\includegraphics[width=\linewidth]{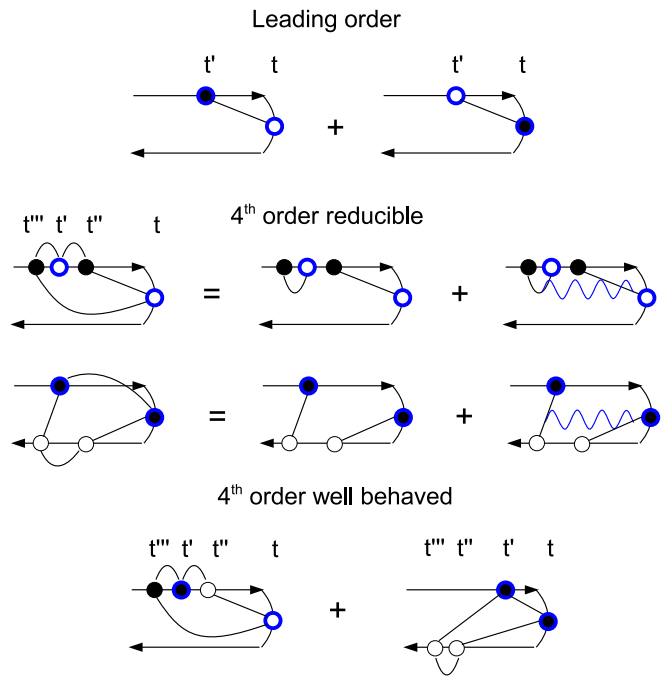}
\caption{(Top) The two diagrams that contribute to current correlations in the leading order.
The left-hand side diagram gives the contribution from forward Cooper-pair tunnelling and the right-hand side diagram
from backward tunnelling. The connection lines visualize pair correlations after taking the ensemble average.
(Middle) Two diagrams contributing to the current in the fourth order. These diagrams can be reduced to a product of two first-order ones
and a term decribing interaction between the pairs (wiggly lines). This is done to pick out the finite asymtotic behaviour for large $t-t'$,
since here the correlator does not decay to zero for finite $t-t''$ and $t'-t'''$ (with increasing $t-t'$).
(Bottom) These fourth-order diagrams decay to zero when $t-t'$ increases and are not needed to be split into two.
}
\label{fig:Keldysh1}
\end{figure}

In Fig.~\ref{fig:Keldysh1}, we show certain diagrams that correspond to calculating the correlator~(\ref{eq:CorrelatorDefinition}). 
We assume $t>t'$, so each diagram ends to a point at time $t$,
and other points occur always before this. The upper branch corresponds to normal time evolution (operators $\hat U$),
whereas the lower branch to reversed time evolution (operator $\hat U^{\dagger}$).
Here we assign a white/black dot to each term $e^{i\omega_{\rm J}x-i\hat \phi_0(x)}$/$e^{-i\omega_{\rm J}x+i\hat \phi_0(x)}$,
corresponding to Cooper-pair tunneling in two directions, where $x$ is the time of the tunneling.
These terms can originate either from the Josephson-energy operators,
$E_{\rm J}\cos\left[ \omega_{\rm J}x -\hat\phi_0(x) \right]=(E_{\rm J}/2)\left[ e^{i\omega_{\rm J}x-i\hat \phi_0(x)}+e^{-i\omega_{\rm J}x+i\hat \phi_0(x)}  \right]$,
or from the two (similarly expanded) junction-current operators, $\hat I^0_{\rm J}(t)$ and $\hat I^0_{\rm J}(t')$.
The ones from current operators are static and are marked as additional blue circles;
the inside can be white or black.
The ones originating from Josephson-energy operators are freely moving, and integration over the corresponding times is performed.
All blue points with white (black) inside give a factor $(-)I_{\rm c}/2i$.
All other points on the upper (lower) Keldysh branch contribute with a factor $(-)iE_{\rm J}/2\hbar$.
The ordering of the (multiplied) operators is indicated by the Keldysh time-arrows: upper branch is stacked (multiplied)
first with normal time ordering, ending current operator at time $t$,
after which one continues with the lower branch operators, multiplied in the opposite time ordering~\cite{Michael2015}.

\begin{figure}[bt]
\includegraphics[width=\linewidth]{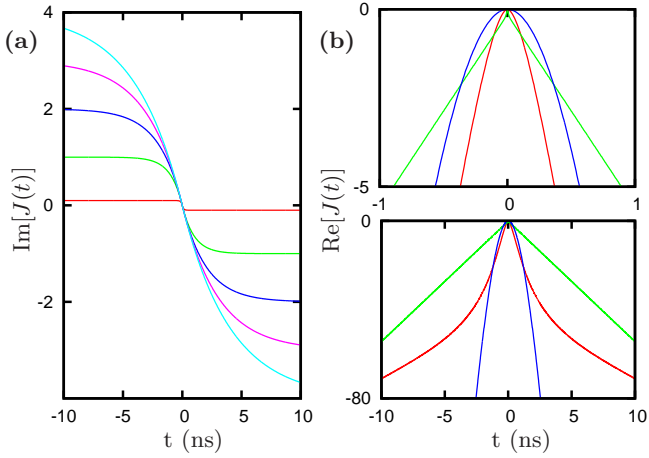}
\caption{
(a) The imaginary part of $J(t)$ for an Ohmic transmission line with $R_{\rm Q}C=1$~ns
and $\pi Z_0/R_{\rm Q}=0.1,1,2,3,4$ (from the bottom to top for $t<0$).
The linear decrease nearby $t=0$ is connected to the finite charging energy $e^2/2C$ and 
is important if its decay, on timescale $Z_0C$, is slow compared to the induced oscillations, 
leading to a cross over at $Z_0\sim R_{\rm Q}$.
(b) The form of the real part is determined by the relative magnitude of the cut-off frequency $\omega_{\rm c}=1/Z_0C$ and thermal
energy $\omega_{T}=k_{\rm B}T/h$. For a typical low-Ohmic case, $\omega_{\rm c}$ dominates
and the function is practically linear (green line, $Z_0/R_{\rm Q}=0.05$, $T=100$~mK, $C=10$~fF).
At short times (top) a transient behaviour occurs,
similar as  at larger times (bottom) for a high-Ohmic environment at zero temperature (red line, $Z_0/R_{\rm Q}=4$, $C=10$~fF).
When the temperature is higher than the energy cut-off, the short-time behaviour is quadratic
(blue line, $Z_0/R_{\rm Q}=4$, $T=20$~mK, $C=100$~fF).
}
\label{fig:CorrelationFunctions}
\end{figure}

\subsubsection{Contraction to pairs and coupled pairs}
In the next step we contract the expectation values of the multiplied Cooper-pair-tunnelings, $e^{\pm i\hat \phi_0(x)}$, into pair correlations.
Only the combinations with equal amount of phases with plus-sign and minus-sign contribute.
This means that there has to be equal amount of black and white points in each diagram we include.
For the leading (second) order diagrams we can use the result
\begin{eqnarray}\label{eq:Contraction2nd}
\left\langle e^{i\hat\phi_0(t)} e^{-i\hat\phi_0(t')} \right\rangle = \left\langle e^{-i\hat\phi_0(t)} e^{+i\hat\phi_0(t')} \right\rangle = e^{J(t-t')}.
\end{eqnarray}
The contraction is expressed by the phase-correlation function
\begin{equation}\label{eq:JofTime}
J(t) = \left\langle \left[\hat \phi_0(t)-\hat \phi_0(0)\right]\hat \phi_0(0)\right\rangle.
\end{equation}
A direct calculation using Eq.~(\ref{PhaseDifference0}) gives
\begin{eqnarray}\label{eq:PhaseCorrelations}
\left\langle\hat \phi_0(t)\hat \phi_0(t')\right\rangle= 2\int_{-\infty}^{\infty} \frac{d\omega}{\omega}  \frac{ {\rm Re}[Z_{\rm t}(\omega)] }{R_{\rm Q}}\frac{e^{-i\omega(t-t')}}{1-e^{-\beta\hbar\omega}},
\end{eqnarray}
where $R_{\rm Q}=h/4e^2$.
The usual forms of $J(t)$ we deal with are plotted in Fig.~\ref{fig:CorrelationFunctions}.
We draw a line connecting the white and black point corresponding to multiplication by function~(\ref{eq:Contraction2nd}).
Similarly, the arbitrary order contributions contract to a function of pair correlations with the form
\begin{eqnarray}\label{eq:Contraction4th}
\left\langle\Pi_i \exp\left[in_i\hat \phi_0(t_i)\right] \right\rangle = \exp\left[ -\sum_{i>j}n_in_j J(t_i-t_j)  \right],
\end{eqnarray}
where the integers $n_i$ take values $\pm 1$ with the constraint $\sum_i n_i=0$.
We note that the total correlation function depends on all paired time-differences.
In Fig.~\ref{fig:Keldysh1}, to visualize this we draw a line that connects all the points.

\subsubsection{Separating asymptotic pair correlations}\label{sec:SeparatingToPairs}
Some fourth order diagrams do not approach zero when
certain time differences approach infinity. This is the case
when white and black points can be grouped to pairs in a such way that their partner will always be a neighboring point on the real time axis.
This is the case for the middle diagrams in Fig.~\ref{fig:Keldysh1}. In this case it turns out to be convenient
(and in the case of photon correlators necessary) to analytically take out the asymptotic behaviour.
We do this by the following trick (corresponding to the upper fourth-order reducible diagram in Fig.~\ref{fig:Keldysh1})
\begin{eqnarray}
&&\left\langle e^{-i\hat\phi_0(t)}e^{i\hat\phi_0(t'')}e^{-i\hat\phi_0(t')} e^{i\hat\phi_0(t''')}\right\rangle=\nonumber \\
&&e^{J(t-t'')+J(t'-t''')}+ e^{J(t-t'')+J(t'-t''')} \times \nonumber\\
&&\times \left[ e^{J(t-t''')+J(t''-t')-J(t-t')-J(t''-t''')} -1 \right]. \nonumber
\end{eqnarray}
The first term on the right-hand side is now the product of bare pair correlations,
whereas the term inside the parentheses describes interaction between these pairs.
The latter goes exponentially to zero when $t-t'$ goes to infinity (assuming a finite temperature, $T>0$).
This is because asymptotically $J(t)=-a\vert t\vert -ib\ {\rm Sgn}(t)+{\rm constant}$, where $a,b>0$.
The fact that the first term  remains for large $t-t'$, describes decoupling of the
current correlator for large $t-t'$,
\begin{equation}\label{eq:disentangling}
\lim_{\vert t-t'\vert \rightarrow\infty}\left\langle \hat I_{\rm J}(t')\hat I_{\rm J}(t) \right\rangle = \left\langle \hat I_{\rm J}(t')\right\rangle \left\langle  \hat I_{\rm J}(t) \right\rangle.
\end{equation}

Below we apply this expansion to investigate the power-spectral densities up to the fourth order in the tunnelling coupling $E_{\rm J}$.

\subsection{Second-order results: Independent charge tunneling}\label{sec:IndependentTunneling}
We first discuss the main results obtained from the leading-order calculation, derived partly in Refs.~[\onlinecite{Leppakangas2013,Leppakangas2014,Portier2}].
This models emission properties in the limit of independent charge tunneling.
We also discuss the important low- and high-Ohmic limits, which provide clear physical interpretations to the derived formulas.

\subsubsection{Current fluctuations for arbitrary electromagnetic environment}
For the correlations of the current fluctuations across the Josephson junction
we obtain 
\begin{eqnarray}\label{eq:CurrentCorrelationLeadingOrder}
\left\langle \hat I_{\rm J}(t) \hat I_{\rm J}(0)  \right\rangle_{\rm 2nd}=\frac{I_{\rm c}^2}{4} \left[ e^{J(t)+i\omega_{\rm J}t} + e^{J(t)-i\omega_{\rm J}t} \right].
\end{eqnarray}
The phase correlation function $J(t)$ is defined in Eq.~(\ref{eq:JofTime}). 
The Fourier transformation of Eq.~(\ref{eq:CurrentCorrelationLeadingOrder}) gives the
power spectral density~\cite{Portier2}
\begin{eqnarray}\label{eq:FiniteFrequency2nd}
&&\left\langle \hat I_{\rm J}(t) \hat I_{\rm J}(0)  \right\rangle_{\omega, {\ \rm 2nd}}\nonumber\\
&=&\frac{\pi\hbar I_{\rm c}^2}{2}\left[ P(\hbar\omega_{\rm J}-\hbar\omega)+P(-\hbar\omega_{\rm J}-\hbar\omega)\right],
\end{eqnarray}
where we have introduced the probability density~\cite{Devoret1990,Ingold1992}
\begin{equation}\label{eq:PEFunction}
P(E)=\int_{-\infty}^{\infty} dt \frac{1}{2\pi\hbar} e^{J(t)}e^{i\frac{E}{\hbar}t}.
\end{equation}
This function gives the probability of the electromagnetic environment to absorb (or
emit if negative) the energy $E$ in  a photon-assisted tunneling event.
Note that the impedance of the environment is not explicit in result~(\ref{eq:FiniteFrequency2nd}),
as it is (as a front factor) in the corresponding photon-flux density, Eqs.~(\ref{eq:CurrentFluctuations}-\ref{eq:Impedance}),
but it influences the form of $P(E)$ through the correlator $J(t)$, see Eqs.~(\ref{eq:JofTime}-\ref{eq:PhaseCorrelations}).

Particularly, at zero-frequency ($\omega=0$) we obtain
\begin{eqnarray}\label{eq:ShotNoiseLeadingOrder}
\left\langle \hat I_{\rm J}(t) \hat I_{\rm J}(0) \right\rangle_{\omega=0, {\rm 2nd}}=2e\left[\left\langle \hat I_{\rm J}^+ \right\rangle_{\rm 2nd} +\left\langle \hat I_{\rm J}^- \right\rangle_{\rm 2nd} \right],
\end{eqnarray}
where we have used the leading-order results for the forward (+) and backward (-) Cooper-pair tunneling,
\begin{eqnarray}\label{eq:LeadingCurrent}
\left\langle \hat I_{\rm J}^{\pm} \right\rangle_{\rm2nd}\equiv I_2^{\pm}=\frac{\pi\hbar I_{\rm c}^2}{4e}P(\pm 2eV).
\end{eqnarray}
This implies that the charge transport is characterized by a Cooper-pair shot-noise.
This noise can equivalently be interpreted to be a result of
vacuum fluctuations of the electromagnetic radiation~\cite{Koch1,Koch2,Likharev}. 



\subsubsection{Low- and high-Ohmic transmission lines}\label{sec:LowOhmicEE}
The interaction strength between a Cooper-pair tunneling event and single-photon emission
is determined by the relation between the characteristic impedance of the TL and the resistance quantum $R_{\rm Q}$.
For example, in the case of a single-mode environment,  the excitation probability from the ground to the $n$-photon state
induced by a tunneling event is $ p^n e^{-p}/n!$, where $p= 4Z_0/R_{\rm Q}$ (for a $\lambda/4$ standing-wave~\cite{Leppakangas2013}).
Thus, for $Z_0\ll R_{\rm Q}$  the interaction is weak. 
Here, single Cooper-pair tunneling occurs predominatly through single-photon emission.
From other processes, the photon pair production is detectable at frequencies well below the Josephson frequency~\cite{Hofheinz2011,Leppakangas2013,Leppakangas2014}.
For this low-Ohmic TL the long-time behaviour $J(t)=-D\vert t\vert$ where $D=2\pi Z_0/R_{\rm Q}\beta \hbar$ is often a sufficient approximation.
Therefore, current correlations [Eq.~(\ref{eq:CurrentCorrelationLeadingOrder})]  are dominated by damped oscillations at frequencies $\pm\omega_{\rm J}$.


For a high-Ohmic TL, $Z_0\gg R_{\rm Q}$, the interaction between charge tunneling and photon emission is strong.
This favours simultaneous emission of a large number of low-frequency photons.
When the cut-off $1/Z_0C$ is the lowest frequency scale ($1/Z_0C<1/\beta h$),
the short-time behaviour of the phase correlator can be approximated as $J(t)=-(\pi/CR_{\rm Q})(it +t^2/\hbar\beta)$.
Such a contribution drops the voltage at the junction by $2e/C$ and
the current correlator oscillates at frequencies $\pm 2eV - 4E_C/\hbar$, where $E_C=e^2/2C$.
The fact that $E_C$ comes into the game reflects a change in the charge transport to a regime
where the tunneled charge first goes to the junction capacitor, and is from there slowly released in the $Z_0C$ timescale.

\subsubsection{Correlated photon flux density and junction current}
In certain limits we expect clear correlations between high-frequency photon emission and low-frequency current fluctuations.
To study this in more detail we can evaluate the correlator between the photon flux density and the junction current.
This correlator is closely related to the third moment of the transferred charge~\cite{Reulet2003,ThirdCumulant}.
We obtain
\begin{eqnarray}\label{eq:CurrentFluxCorrelator}
&&\left\langle \hat a^{\dagger}_{\rm out}(\omega_1)\hat a_{\rm out}(\omega_2)  \hat I_{\rm J}(\omega_3) \right\rangle = \\
&&\frac{2\pi eI_{\rm c}^2Z_0A^*(\omega_1)A(\omega_2)}{\sqrt{\omega_1\omega_2}}P(\hbar\omega_{\rm J}-\hbar\omega_1) \delta(\omega_1-\omega_2-\omega_3),  \nonumber
\end{eqnarray}
where we use the notation $ \hat I(\omega_3)=\int_{-\infty}^{\infty} dt e^{ i\omega_3 t} \hat I_{\rm J}(t)$.
We consider a perturbative expansion up to the second order in the tunneling coupling and
neglected contributions proportional to the Bose factor within assumption $\omega_1\gg k_{\rm B}T/\hbar $.
If we now consider the photon flux nearby a certain frequency $\omega_0$ within a bandwidth BW$\ll\omega_0$,
which means $\omega_1,\omega_2\in (\omega_0-{\rm BW},\omega_0+{\rm BW})$, it follows that
the delta-function in Eq.~(\ref{eq:CurrentFluxCorrelator}) forces  the current operator $\hat I_{\rm J}(\omega_3)$ to be in the
low-frequency domain $\omega_3< 2\ {\rm BW}$.
Particularly, for $\omega_3=0$ we  obtain
\begin{eqnarray}
\int \frac{d\omega_2}{2\pi}\left\langle  \hat a^{\dagger}_{\rm out}(\omega_1)\hat a_{\rm out}(\omega_2)  \ \hat I_{\rm J}(\omega_3=0)\right\rangle = 2e f(\omega).
\end{eqnarray}
Here $f(\omega)$ is the leading-order result for the flux density of the emitted photons, determined by Eqs.~(\ref{eq:CurrentFluctuations}) and~(\ref{eq:FiniteFrequency2nd}).
This describes perfectly correlated high-frequency photon emission with low-frequency current fluctuations.

\subsection{Fourth-order results: Correlations between tunneling events}\label{sec:DependentTunneling}
To study the first corrections from higher orders,
we expand the current-current correlator as
\begin{eqnarray}\label{eq:4thOrderCorrelations}
&&\left\langle \delta \hat I_{\rm J}(t)\delta \hat I_{\rm J}(0) \right\rangle\equiv \left\langle \left[ \hat I_{\rm J}(t)-I \right]\left[ \hat I_{\rm J}(0)-I \right] \right\rangle \nonumber \\
&&\approx\left\langle\hat I_{\rm J}(t)\hat I_{\rm J}(0) \right\rangle_{\rm 2nd} + \left[ \left\langle \hat I_{\rm J}(t)\hat I_{\rm J}(0) \right\rangle_{\rm 4th} - I_2^2 \right].
\end{eqnarray}
The first term at the bottom line was evaluated in the preceding section, see Eq.~(\ref{eq:CurrentCorrelationLeadingOrder}).
In the following we concentrate on the term inside the square brackets.
This term approches zero within the
memory time of the electromagnetic environment [following from Eq.~(\ref{eq:disentangling})].
The memory time here is the maximum of the inverse dephasing rate and the inverse frequency cut-off.
Therefore, the behaviour at short times describes nonequilibrium between consecutively tunneling Cooper pairs.
Generally, this type of correlations can be measured by investigating changes in the emission spectrum with increasing $E_{\rm J}$.

\begin{figure}[tb]
\includegraphics[width=0.9\linewidth]{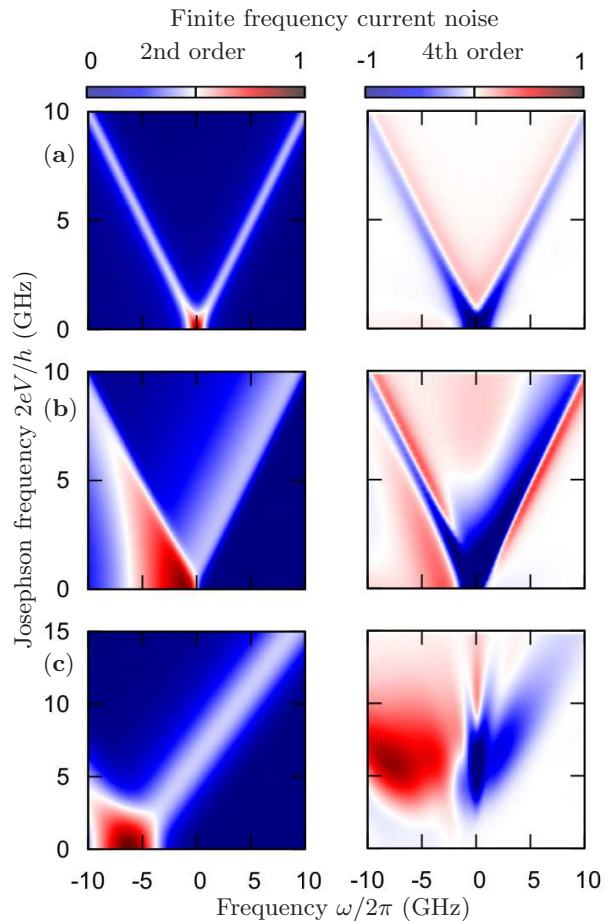}
\caption{
Spectral density of the current fluctuations in the second- (left) and in the fourth-order (right)
for three Ohmic transmission lines (TL) as a function of the bias voltage $V$ (a.u.).
The positive frequencies describe emission and the negative absorption.
(a) For the low-Ohmic TL ($Z_0/R_{\rm Q}=0.05$, $T=100$~mK) 
most of the fluctuations appear at the Josephson freqeuency $2eV/h$, seen as the V-shaped resonance lines.
The fourth-order contribution describes their weak redirection to lower frequencies.
(b) For increased TL resistance ($Z_0/R_{\rm Q}=0.5$, $T=10$~mK)  lower-frequency fluctuations increase,
due to the increased probability for multi-photon production during CP tunneling.
The fourth-order spectral density mainly redirects fluctuations to lower frequencies.
(c) For the high-Ohmic TL  ($Z_0/R_{\rm Q}=4$, $T=10$~mK)
the second-order fluctuations are shifted to lower frequencies, $\pm\omega_{\rm J}-2e^2/C\hbar$, corresponding to the energy loss when charging the junction capacitor by a Cooper pair.
The fourth-order spectral density describes redirection of the fluctuations to lower frequencies but is also
characterized by zero-frequency dip or peak, corresponding to blockade or enhancement of consecutive Cooper-pair tunneling.
For all plots $C=10$~fF.
}
\label{fig:resultsII}
\end{figure}

\subsubsection{Numerical results: From low- to high-Ohmic transmission lines}
Numerical results for the finite-frequency current-noise in the second and in the fourth order of tunneling coupling $E_{\rm J}$ are shown
in Fig.~\ref{fig:resultsII}.
We study here the change in the fluctuation spectrum
when  increasing $Z_0$ from well below to beyond $R_{\rm Q}$. 
As discussed above, in the  regime $Z_0\ll R_{\rm Q}$,
the most junction-current fluctuations occur at the Josephson frequency $\omega=\pm\omega_{\rm J}$,
seen as the V-shaped resonance lines in the second-order results, left-hand side of Fig.~\ref{fig:resultsII}(a).
The fourth-order spectrum corrects this result. The main contribution is again at the Josephson frequency, but
it is negative just above $\omega_{\rm J}$ and positive just below, see the right-hand side of Fig.~\ref{fig:resultsII}(a).
Therefore, it redirects fluctuations from $\omega_{\rm J}$ to a slighty lower frequency.
This is consistent with a reduction of the junction voltage due to the net current across the system, Eq.~(\ref{eq:VoltageReduction}). 
The change this contribution can describe is limited, as too high $E_{\rm J}$ takes the total spectrum to negative values.
We also note that the zero-frequency noise increases/decreases for high/low bias voltages, consistent with results presented in Ref.~[\onlinecite{Ingold2002}]. There, a disappearance of the noise at low voltages (or high $E_{\rm J}$) is
derived, reflecting trapping of the phase difference by the Josephson potential energy, supporting noiseless supercurrent.
As we are here calculating small corrections to the leading-order results, our low-voltage result
can be interpreted as a first sign of this effect (for small $E_{\rm J}$).
Other types of correlations between Cooper-pair tunneling events are weak.

With increasing characteristic impedance,
more fluctuations appear at frequencies below $\omega_{\rm J}$, see Fig.~\ref{fig:resultsII}(b).
This is due to the increased probability for multi-photon assisted Cooper-pair tunneling.
Here, the corrections from the fourth-order  reduce radiation in the neighbourhood of the now wider main peak and redirects it to lower frequencies.
This can be understood as stronger voltage reduction after each tunneling.
Interestingly, there is also enhancement of radiation above $\omega=\omega_{\rm J}$ (and also nearby $\omega>-\omega_{\rm J}$). Here, the fluctuations
are missing in the leading-order results, as they would need the aid of incoming high-frequency thermal radiation.
The presence of them in the fourth-order contribution hints towards a process,
where two Cooper-pairs tunnel across the junction and in total emit two photons that satisfy $\omega_1+\omega_2=2\omega_{\rm J}$.

In the limit $Z_0\gg R_{\rm Q}$,
the charging energy of a single Cooper pair ($4E_C$) plays an essential role. 
Here, the tunneled charge goes first to the junction capacitor,
and is from there released in the timescale $Z_0C$.
This picture suggests a large change in the junction voltage after each Cooper-pair tunneling,
and strong correlations between consecutively tunneling charges.
In Fig.~\ref{fig:resultsII}(c) we see that the fourth-order contribution indeed
brings in reduction of fluctuations at the frequency $(2eV-4E_C)/\hbar$. 
An interesting feature is the clear zero-frequency dip or peak, whose width can be traced to be $\sim 1/Z_0C$.
It describes temporal changes in the low-frequency current noise due to a blockade or enhancement
of consecutive Cooper-pair tunneling during the $Z_0C$-recovery of the junction voltage.
We investigate this feature further below.

\subsubsection{Numerical results: Second-order coherence of Cooper-pair transport}\label{sec:G2CP}
For further illustration of the low-frequency effects in the case $Z_0\gg R_{\rm Q}$ we plot the correlations in real time.
In the considered case there is a drastic difference between junction current fluctuations and junction voltage fluctuations,
due to the strong shunting of the parallel capacitor (cut-off frequency $1/Z_0C\approx 2\pi\times 0.6$~GHz).
This means that high-frequency oscillations become hardly observable for outside detection.
We then study the function
\begin{equation}\label{eq:SecondCoherenceVoltage}
g^{(2)}_{\rm CP}(t)\equiv  {\rm Re}\left[\int_{-\infty}^{\infty} dt' {A}(t-t')  \frac{\left\langle \hat I_{\rm J}(t')\hat I_{\rm J}(0)  \right\rangle_{\rm 4th}}{I_2^2}\right] -1,
\end{equation}
where the convolution through ${A}(t)$ stands for filtering of the correlation function by $\vert A(\omega)\vert^2$.
We include contribution only from positive frequencies, that means only from emission.
We name this the second-order coherence of Cooper-pair transport, in analogy to the definition introduced in Ref.~[\onlinecite{GermanGuys}].
This function does not depend on $E_{\rm J}$.
The total time integral $\int_0^{\infty} dt g^{(2)}_{\rm CP}(t)$ describes increase $(>0)$
or reduction $(<0)$ of the zero-frequency shot noise due to non-equilibrium
between consecutively tunneling Cooper pairs.
The momentary negativity (positivity) hints towards temporal blockade (enhancement) of consecutive Cooper-pair tunneling.
For long time separations the current correlations vanish and we have $\lim_{t\rightarrow \infty} g^{(2)}_{\rm CP}(t)=0$.

\begin{figure}[tb]
\includegraphics[width=\linewidth]{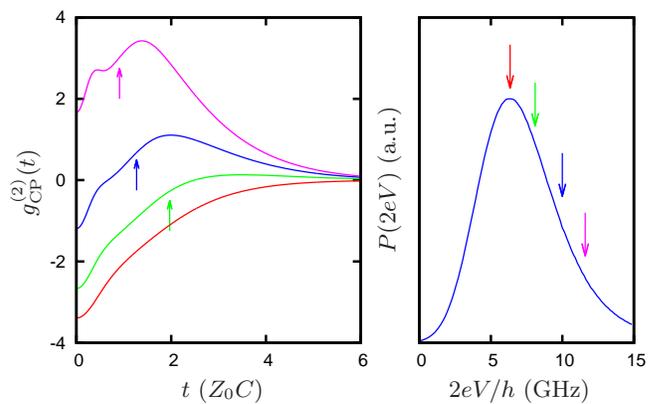}
\caption{
(Left) The second-order coherence of Cooper-pair transport as defined in Eq.~(\ref{eq:SecondCoherenceVoltage})
for the setup of Fig.~\ref{fig:resultsII}(c).
(Right) The corresponding $P(E)$-function and the studied voltage-bias points (arrows).
The $g^{(2)}_{\rm CP}(t)$-function describes temporal correlations in the junction-current fluctuations
originating in nonequilibrium between consecutively tunneling Cooper pairs.
Coulomb blockade during recharging of the junction reduces the zero-frequency noise ($\int dt g^{(2)}_{\rm CP}(t)<0$)
whereas enhanced Cooper-pair tunneling rate increases the noise ($\int dt g^{(2)}_{\rm CP}(t)>0$).
At long time separations the correlations vanish, $\lim_{t\rightarrow \infty} g^{(2)}_{\rm CP}(t)=0$.
}
\label{fig:resultsCPG2}
\end{figure}

In Fig.~\ref{fig:resultsCPG2}, we plot numerical results for $g^{(2)}_{\rm CP}(t)$
in the setup of Fig.~\ref{fig:resultsII}(c) for several values of the bias voltage $V$.
At resonance $2eV=4E_C$ [red arrow in Fig.~\ref{fig:resultsCPG2}(b)] 
the $g^{(2)}_{\rm CP}(t)$ function is negative  and decays towards zero in the timescale given by $Z_0C$.
This behaviour is consistent with a semiclassical interpretation,
where the first tunneling drops the junction voltage by $2e/C$ and results in a momentary blockade
of further Cooper-pair tunneling events.
Here, we expect the junction voltage recovery to follow the formula $V-(2e/C)e^{-t/Z_0C}$.
For higher bias voltages the negativity decreases and creates a maximum at certain time $t>0$. 
This is understood as that here the recovering junction voltage sweeps through
a value that provides a higher (consecutive) Cooper-pair tunneling rate.
In the semiclassical picture this occurs at times pointed by the corresponding arrows in Fig.~\ref{fig:resultsII}(a).
At higher voltages the $g^{(2)}_{\rm CP}(t)$-function is also superposed by low-frequency oscillations,
describing the absence or increase of low-frequency radiation at specific frequencies.
These are identified as the local minima or maxima in the right-hand side of Fig.~\ref{fig:resultsII}(c).

\subsection{Convergence}
In this and in the next section we work in the perturbative limit and
implicitly assume weak tunneling rates compared to the relaxation rates in the electromagnetic environment.
Generally, we are pleased to see that the fourth-order expressions converge, that is, are not infinite.
Therefore, the results are always sound, for small enough $E_{\rm J}$. What is small depends on the details of the biasing circuit.

In the case of a low-Ohmic TL we demand that the phase noise from the leading-order results should
at all frequencies be smaller than from the zeroth-order contribution.
For high bias voltages this leads to the condition~\cite{Leppakangas2014}
\begin{equation}
\frac{E_{\rm J}^2}{2eV} \ll 4k_{\rm B}T.
\end{equation}
This can equivalently be interpreted as a demand that the thermal dephasing of the phase difference across the Josephson junction has to be
faster than the rate for inelastic Cooper-pair tunneling.

For a high-Ohmic environment
this condition is rather strict since the $P(E)$-function is well behaved also at zero temperature (has no singularity).
We can apply here a condition known from the $P(E)$-theory~\cite{Ingold1992},
$E_{\rm J}P(E)\ll 1$.
The analysis made here makes it explicit that we need weak tunneling rates for all possible
momentary values of the junction voltage $V_{\rm J}$,
that is for all $E$.
At zero temperature the $P(E)$-function has a maximum at $E=4E_C$ and is broadened by $1/Z_0C$.
This implies the condition
\begin{equation}\label{eq:HORelation1}
E_{\rm J}\ll \frac{h}{Z_0C}.
\end{equation}
For temperatures much higher than the frequency cut-off
one replaces $ h/Z_0C$ by the corresponding broadening of the $P(E)$-function. This leads to the condition
\begin{equation}\label{eq:HORelation2}
E_{\rm J}\ll\sqrt{4\pi k_{\rm B}TE_C}.
\end{equation}
The constrictions~(\ref{eq:HORelation1}-\ref{eq:HORelation2}) can also be derived from a more general convergence analysis on the
Keldysh contour~\cite{Michael2015}.

\section{Photon bunching}\label{sec:Photons}
We now turn our attention to statistics of the emitted photons. 
In the preceding sections we learned that the radiation power is given,
up to a frequency-dependent front-factor and a term describing thermal noise, by the
finite-frequency current noise at the Josephson junction. However, the general moments
of the junction current and that of the emitted field are not equivalent, even though they can
 be mapped to each other in a nontrivial way, as discussed below.
The difference stems from the fact that a Cooper-pair tunneling can occur with an arbitrary multi-photon emission.
Still, direct similarities are expected, especially in situations where the relevant physics is dominated by single-photon-assisted
Cooper-pair tunneling. We study this explicitly in the cases of low- and (stepped) high-Ohmic transmission lines.

\subsection{General definitions}
We concentrate on statistics of observing two photons of the same frequency within a time interval $\tau$
by evaluating the so-called second-order coherence.
This is a standard tool for higher-order characterization
of the electromagnetic radiation~\cite{WallsMilburn}. For a single-mode electromagnetic field it is defined as
\begin{equation}\label{eq:SingleG2}
g_{\rm single}^{(2)}(\tau)=\frac{\left\langle \hat a^{\dagger} \hat a^{\dagger}(\tau)  \hat a(\tau) \hat a \right\rangle}{\left\langle \hat a^{\dagger} \hat a  \right\rangle^2  }.
\end{equation}
This can be interpreted as the probability to detect a photon at time $t=0$ and a second photon at time $t=\tau$,
normalized by the probability of two uncorrelated detections.
Coherent radiation is characterized by Poissonian statistics, $g^{(2)}(0)=1$, whereas antibunched photons by $g^{(2)}(0)<1$, and
bunched photons by $g^{(2)}(0)>1$.
Thermal radiation, or generally radiation with no phase memory, appears to an observer bunched, $g^{(2)}(0)=2$.

For a continuous-mode field the second-order coherence and its interpretations
are analogous~\cite{Loudon}. First we define the (unnormalized) second-order coherence,
\begin{eqnarray}\label{eq:SecondCoherence}
&&G^{(2)}(\tau)\equiv \left( \frac{\hbar Z_0}{4\pi} \right)^2 \int_{\rm BW} \sqrt{\omega_1\omega_2\omega_3\omega_4}   \\
&\times & e^{i\tau(\omega_2-\omega_3)}\left\langle \hat a_{\rm out}^{\dagger}(\omega_1)\hat  a_{\rm out}^{\dagger}(\omega_2)\hat  a_{\rm out}(\omega_3) \hat a_{\rm out}(\omega_4) \right\rangle.\nonumber
\end{eqnarray}
Here BW stands for the frequency range we integrate over, i.e.~the bandwidth of the detector.
This can be described by the band-pass filter $F(\omega)$,
\begin{eqnarray}
\int_{\rm BW}&\equiv& \int_{0}^{\infty}d\omega_1 \int_{0}^{\infty} d\omega_2 \int_{0}^{\infty} d\omega_3  \int_{0}^{\infty} d\omega_4\nonumber\\
&\times& F(\omega_1)F(\omega_2)F(\omega_3)F(\omega_4).\nonumber
\end{eqnarray}
In this article we consider a Gaussian filter centered at certain measurement frequency $\omega_{\rm m}$,
\begin{equation}
F(\omega)=\exp\left[ -\frac{(\omega-\omega_{\rm m})^2}{2\Delta^2} \right].
\end{equation}
We have therefore $\int d\omega F(\omega)=\sqrt{2\pi}\Delta$.
The corresponding continuous-mode first-order coherence has the form,
\begin{eqnarray}\label{eq:FirstOrderCoherence}
G^{(1)}(\tau)\equiv \frac{\hbar Z_0}{4\pi}\int_{\rm BW}\sqrt{\omega_1\omega_2}\left\langle \hat a_{\rm out}^\dagger(\omega_1) \hat a_{\rm out}(\omega_2)  \right\rangle e^{i\omega_1\tau}, \nonumber
\end{eqnarray}
where $\int_{\rm BW}$ is defined analogously.
The first-order coherence is a Fourier transform of the emission power spectrum.
In Section~\ref{sec:power}
this was evaluated (within neglecting thermal fluctuations in the transmission line) up to fourth order in the critical current.

The normalized second-order coherence of a continuous-mode field
is defined as,
\begin{equation}\label{eq:G2ContinuousMode}
g^{(2)}(\tau)=\frac{G^{(2)}(\tau)}{\vert G^{(1)}(0)\vert^2 }.
\end{equation}
Important here is that for a continuous-mode field $g^{(2)}(\tau)$ depends on the frequency range we integrate over,
i.e.~which frequency photons (and at which timescale) we measure.

\subsection{Second-order results: Independent charge tunneling}
We first repeat the main results from the leading-order calculation~\cite{Leppakangas2013,Leppakangas2014}.
Similarly as in the case of current fluctuations at the Josephson junction, this order includes radiation only from a single tunneling event.
Inserting the solution of Eq.~(\ref{eq:OutputSolution}) to the continuous-mode second-order coherence, Eq.~(\ref{eq:SecondCoherence}),
we obtain ($\omega\gg k_{\rm B}T/\hbar$)
\begin{eqnarray}\label{BunchingLeadingOrder} 
G^{(2)}_{\rm 2nd}(\tau) &=& \left( \frac{\hbar Z_0}{4\pi} \right)^2\frac{ 2\pi I_{\rm c}^2}{R_{\rm Q}}\int_{\rm BW} A_\omega e^{i\tau(\omega_2-\omega_3)} \\
&\times&   P[\hbar(\omega_{\rm J}-\omega_1-\omega_2)] \ \delta(\omega_1+\omega_2-\omega_3-\omega_4).\nonumber
\end{eqnarray}
We use the notation
\begin{equation}
A_\omega\equiv A^*(\omega_1) A^*(\omega_2) A(\omega_3) A(\omega_4). \nonumber
\end{equation}
The surprising feature is that the second-order coherence of Eq.~(\ref{BunchingLeadingOrder})
is practically determined by the $P(E)$-function.
It is therefore analogous to the current spectral density, Eq.~(\ref{eq:FiniteFrequency2nd}),
with the difference that here
the argument of $P(E)$ is the sum of the two detection frequencies (instead of a single one).
Physically, this can be interpreted as down-conversion of photons from the Josephson frequency.
The leading-order result, Eq.~(\ref{BunchingLeadingOrder}), describes also nonclassical
cross-correlations between the  down-converted photons~\cite{Leppakangas2013,Leppakangas2014}.


Important in the following is that for the observation of two photons  at the Josephson frequency we have,
\begin{equation}\label{eq:G2Vanishing}
\omega_1=\omega_2=\omega_{\rm J} \ \ \ \rightarrow \ \ \ G^{(2)}_{\rm 2nd}(\tau)\propto P[-\hbar\omega_{\rm J}]\propto e^{-\beta\hbar\omega_{\rm J}}.
\end{equation}
This is negligible at the considered temperatures.
Clearly, emission of two photons should be independent for long time-separations $\tau$,
and the second-order coherence cannot be zero, as implied by Eq.~(\ref{eq:G2Vanishing}).
Below, we find that it is the fourth-order contribution
in the tunneling coupling that is the leading one to account for consecutive single-photon emission.

\subsection{Fourth-order evaluation: Correlations between tunneling events}
We now present evaluation of the second-order coherence up to the fourth order in the tunneling coupling $E_{\rm J}$. To do this
we need to perform an ensemble average
\begin{equation}\label{eq:GeneralCorrelator}
\left\langle \hat a^{\dagger}_{\rm out}\hat a^{\dagger}_{\rm out}\hat a_{\rm out} \hat a_{\rm out} \right\rangle \rightarrow \sum_{i,j,k,l} \left\langle \hat a_i^{\dagger}\hat a_j^{\dagger}\hat a_k\hat a_l \right\rangle,
\end{equation}
with constraint $i+j+k+l=4$.
The indices refer to different-order solutions in $E_{\rm J}$, see Section~\ref{sec:Model}.
The terms that have $i=0$ or $l=0$ describe (inelastic) scattering of incoming thermal radiation.
They are small if $\omega\gg k_{\rm B}T/\hbar$, which we assume to be the case. We then  consider only terms which have $i,l \neq 0$.

The final result can be divided into three type of contributions, 
\begin{eqnarray}\label{eq:G2Division}
G^{(2)}_{{\rm 4th}}(\tau)= \left[{\cal I}_2(\tau) \ + \ {\cal I}_3(\tau) \ + \ {\cal I}_4(\tau)\right] .
\end{eqnarray}
We map the terms either to second, third, or fourth moments of the junction current, i.e.~to
$\left\langle \hat I_{\rm J}^2\right\rangle$, $\left\langle \hat I_{\rm J}^3\right\rangle$,
or $\left\langle \hat I_{\rm J}^4 \right\rangle$, correspondingly.
All contributions are calculated up to the fourth order in $E_{\rm J}$.
For details see Appendix B~and~D.

The term we concentrate on in the following is ${\cal I}_4 (\tau)$, 
\begin{eqnarray}\label{eq:G2MainContribution}
{\cal I}_4(\tau) &=& \left( \frac{Z_0I_{\rm c}}{4\pi} \right)^4 \int_{\rm BW}  e^{i\tau(\omega_2-\omega_3)} A_{\omega}  \\
&\times& \int_{\rm times} \left \langle {\cal T}^{\dagger}\left\{ \hat I^{\dagger}_{\omega_1,t_1} \hat I^{\dagger}_{\omega_2,t_2} \right\}  \  {\cal T}\left\{\hat I_{\omega_3,t_3} \hat I_{\omega_4,t_4}\right\} \right \rangle . \nonumber
\end{eqnarray}
Here we have defined an integration over all tunnelling timings,
\begin{equation}
\int_{\rm times}\equiv\int_{-\infty}^{\infty}dt_1\int_{-\infty}^{\infty}dt_2\int_{-\infty}^{\infty}dt_3\int_{-\infty}^{\infty}dt_4,
\end{equation}
and operators,
\begin{eqnarray}
\hat T(t)&=&\exp\left[ i [ \hat \phi_0(t)- \omega_{\rm J} t]\right]\\
\hat I_{\omega,t}&=& e^{i\omega t}\left[ \hat T^{\dagger}(t) - \hat T(t) \right].
\end{eqnarray}
The operator $\hat T^{(\dagger)}$ describes a creation of a photon of frequency $\omega$ via forward (backward) tunneling event.
The two time-ordering operators $\cal T$ state that the tunneling operators have to be ordered such that
the inner operators in the ensemble average of Eq.~(\ref{eq:G2MainContribution}) occur always timewise later than the neighbouring outer one.

The expression~(\ref{eq:G2MainContribution}) is our central result and is analogous to the single-mode $g^{(2)}(\tau)$, see Eq.~(\ref{eq:SingleG2}).
In Eq.~(\ref{eq:G2MainContribution}) we have a time-ordering of the individual tunneling events, not only of the central times $\tau$.
This is closely related to the causality in the photodetection theory~\cite{Keldysh}.
It has here also a deeper meaning, connecting the normal ordering used in quantum optics for photon counting and the
Keldysh ordering in quantum transport.
It can also be seen as a highly non-trivial check of the correctness of the calculation.
Similar time ordering for photon counting has been used also in other optoelectronical setups~\cite{Beenakker2001}.

\subsection{Fourth-order results: Asymptotic behaviour}
From now on we study situations where the contribution ${\cal I}_4(\tau)$ in Eq.~(\ref{eq:G2Division}) is dominating,
e.g.~low-Ohmic TL and radiation in the neighbourhood of the Josephson frequency.
Similarly as current correlations disentangle at long times, Eq.~(\ref{eq:disentangling}),
the same applies also for the power correlations in the considered second-order coherence. To study this in more detail,
we define now average timings of the outer and inner frequency pairs (pairs $\omega_1 \leftrightarrow \omega_4$ and $\omega_2  \leftrightarrow \omega_3$),
\begin{equation}\label{eq:ContractionTimings}
\tau_1=\frac{t_1+t_4}{2} \,\,\,\, , \,\,\,\, \tau_2=\frac{t_2+t_3}{2}.
\end{equation}
We assume $t_1<t_2$ and $t_3>t_4$, and therefore $\tau_2>\tau_1$. Other orderings are treated similarly.
For timewise well-separated pairs we obtain,
\begin{eqnarray}
&& \lim_{\tau_2 -\tau_1 \rightarrow\infty}  \left \langle {\cal T}^{\dagger}\left\{ \hat I^{\dagger}_{\omega_1,t_1} \hat I^{\dagger}_{\omega_2,t_2} \right\}  \  {\cal T}\left\{\hat I_{\omega_3,t_3} \hat I_{\omega_4,t_4}\right\} \right \rangle=\nonumber \\ 
&& \left\langle   \hat I_{\omega_1,t_1}^{\dagger} \hat I_{\omega_4,t_4} \right\rangle \left\langle I_{\omega_2,t_2}^{\dagger} \hat I_{\omega_3,t_3}  \right\rangle.\label{Eq:ChaoticRelation}
\end{eqnarray}
We note however that this is always non-zero as long as $\vert t_1-t_4\vert$ and $\vert t_2-t_3\vert$ are finite.
In the next step, we subtract this asymptotic contraction analytically and perform the time integrations (for details see Appendix~C),
\begin{eqnarray}\label{eq:SecodOrderDensity}
{\cal I}_4(\tau) &=&\left( \frac{Z_0}{4\pi} \right)^2 \int_{\rm BW}  e^{i\tau(\omega_2-\omega_3)} \left[  s(\omega_1,\omega_4)s(\omega_2,\omega_3) \right.   \nonumber\\
 &+& \left. s(\omega_1,\omega_3)s(\omega_2,\omega_4) + {\cal C}(\omega_1,\omega_2,\omega_3,\omega_4) \right]  .
\end{eqnarray}
Here $s(\omega,\omega')= \hbar\sqrt{\omega\omega'}\left\langle \hat a^{\dagger}_1(\omega) \hat a_1(\omega')   \right\rangle$ is the leading-order result for the emission power density,
\begin{equation}\label{eq:PowerDensityNew}
s(\omega,\omega') = \pi\hbar I_{\rm c}^2{\rm Re}[Z_{\rm t}(\omega)] P(\hbar\omega_{\rm J}-\hbar\omega) \delta(\omega-\omega').
\end{equation}
[Note that $s(\omega)=(1/2\pi)\int d\omega's(\omega,\omega')$.]
The product of two such functions describes indepedent emission with no phase coherence.
The term ${\cal C}(\omega_1,\omega_2,\omega_3,\omega_4)$ accounts for correlations between the consecutively tunneling Cooper pairs.
In the considered cases it is a non-singular function of three independent frequencies.

The integration over the frequencies  ($\int_{\rm BW}$) leads to the result~\cite{Leppakangas20152},
\begin{equation}\label{eq:BunchingMainResult}
g^{(2)}(\tau)= \frac{ {\cal I}_4(\tau) }{\vert G^{(1)}_{\rm 2nd}(0)\vert^2}= 1 + \left\vert \frac{G^{(1)}_{\rm 2nd}(\tau)}{G^{(1)}_{\rm 2nd}(0)} \right\vert^2 + {\cal G}(\tau).
\end{equation}
The second and the third term decay eventually to zero with increasing time, and we have
the property,
\begin{equation}
\lim_{\tau \rightarrow \infty} g^{(2)}(\tau)=1.
\end{equation}
This means Poissonian (independent) photon detection.
The third term on the right-hand side of Eq.~(\ref{eq:BunchingMainResult})
carries information on correlations between consecutively tunneling Cooper pairs,
\begin{eqnarray}\label{eq:bunchingdensity}
{\cal G}(\tau)\equiv \int_{\rm BW} e^{i\tau(\omega_2-\omega_3)}\frac{{\cal C}(\omega_1,\omega_2,\omega_3,\omega_4)}{\vert G^{(1)}_{\rm 2nd}(0)\vert^2}.\nonumber
\end{eqnarray}
We can deduce from Eq.~(\ref{eq:SecodOrderDensity}) that its contribution approaches zero when BW goes to zero. We have then
\begin{equation}
\lim_{{\rm BW} \rightarrow 0} g^{(2)}(0)=2.
\end{equation}
This is a well-known result for chaotic radiation~\cite{WallsMilburn}.
This means that even though photons are emitted independently, the absence of relative phase coherence makes them appear bunched in a detector.
This is consistent with the fact that for infinitesimal bandwidth the time to detect one photon is infinitely long, and
the radiation in our system has only a finite phase-coherence time.

\subsection{Numerical results}\label{sec:NumericalPhotons}

\begin{figure}[tb]
\includegraphics[width=\linewidth]{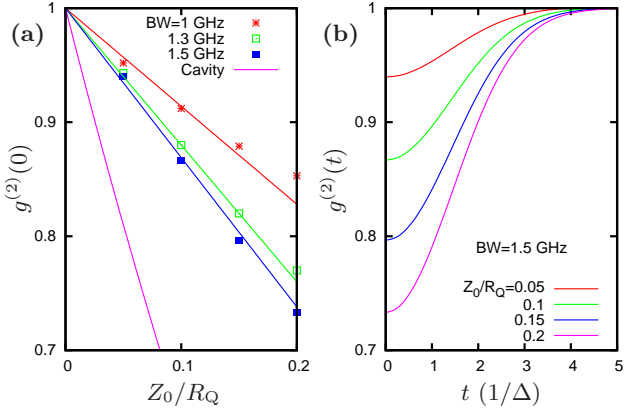}
\caption{
(a)
Second-order coherence $g^{(2)}(0)$ of photons at the Josephson frequency $2eV/\hbar$ for several low-Ohmic transmission lines
and detection bandwidths.
In the limit $Z_0\ll R_{\rm Q}$ the statistics are close to Poissonian, $g^{(2)}(0)\approx 1$.
The second-order coherence decreases approximately linearly with increasing $Z_0/R_{\rm Q}$, meaning that photons become increasingly antibunched.
Similar behaviour is also found in the case of a single-mode cavity~\cite{Ulm2013} (pink line).
Here ${\rm BW}=\Delta/\pi$.
(b) The time dependence $g^{(2)}(t)$ for the largest bandwidth in (a).
The antibunching vanishes in the timescale of the detector sensitivity $1/\Delta$.
The parameters used are $\omega_{\rm J}/2\pi=3.7$~GHz, $T=10$~mK, and $C=10$~fF.
To obtain numerical convergence we have used an additional sharp cut-off of environmental modes above~$13$~GHz.
}
\label{fig:resultsG2LO}
\end{figure}

In Fig.~\ref{fig:resultsG2LO},
we plot the second-order coherence
of photons at the Josephson frequency for an Ohmic TL.
The results are based on numerical evaluation of Eq.~(\ref{eq:BunchingMainResult}).
In the limit $Z_0/R_{\rm Q}\rightarrow 0$
the "correlation density" ${\cal C}(\omega_1,\omega_2,\omega_3,\omega_4)$ in Eq.~(\ref{eq:SecodOrderDensity})
is negative and centered around the Josephson frequency
with an approximative total volume $-1$ and a characteristic width $D=2\pi Z_0/R_{\rm Q}\beta\hbar$.
This means that the photons appear independent if the detection time is much faster than thermal dephasing, $g^{(2)}(0)\approx 1$.
When increasing $Z_0/R_{\rm Q}$, it becomes clear
that at low temperatures the emitted photons near the Josephson frequency are actually antibunched, $g^{(2)}(0)<1$.
In Fig.~\ref{fig:resultsG2LO}(a), we see that the antibunching deepens approximately linearly with $Z_0/R_{\rm Q}\ll 1$.
This observation is similar to earlier predictions for a resonantly driven single mode environment~\cite{Ulm2013}, where
 $g^{(2)}(0)=(1-\kappa/2)^2$. For a $\lambda/4$-type mode as in Ref.~[\onlinecite{Leppakangas2013}] we would have $\kappa= 4Z_0/R_{\rm Q}$,
 plotted as the pink line in Fig.~\ref{fig:resultsG2LO}(a).
The numerical results then indicate that for a "flat" (Ohmic) mode spectrum the antibunching emerges slower with increasing $Z_0/R_{\rm Q}$.
It also depends on the measurement bandwidth. 

In Fig.~\ref{fig:resultsG2LO}(b), we plot the time dependence of the second-order coherence.
In the considered situation the correlations decay in the timescale of the detector sensitivity, $1/\Delta$.
The inverse frequency cut-off ($Z_0C$) is the fastest timescale and related effects are not observed.
In additional simulations we also find that when recording photons 
only well above the Josephson frequency $\omega_{\rm J}$ the photons are highly bunched, $g^{(2)}\gg 2$.
Here the two-photon emission is understood to be rather a repeated single photon emission process,
triggered by  an occasional large fluctuation of the junction voltage.
Furthermore, photons well below $\omega_{\rm J}$ have $g^{(2)}(0)\approx 2$.
Here, most of the detected photons are one half of an emitted pair, whose individual frequencies sum to $\omega_{\rm J}$.
Such photon pair production is triggered by vacuum fluctuations of the electromagnetic environment~\cite{Leppakangas2013}.

We can compare these results to Cooper-pair transport statistics.
In Ref.~[\onlinecite{Ingold2002}] the current noise  has been evaluated to all orders in $E_{\rm J}$
for the specific case of zero temperature and high cut-off frequency.
It is found that for a high voltage bias the Cooper-pair Fano factor is (slightly higher but) very close to one.
This is consistent with the fact that the corresponding $g^{(2)}_{\rm CP}(t)$-function of Cooper pairs (not plotted) practically shows  only decaying oscillations at $\omega_{\rm J}$
with a small but positive weight $\int_0^\infty dt g^{(2)}_{\rm CP}(t)\gtrsim 0$. This behaviour can also be deduced from the right-hand side of Fig.~\ref{fig:resultsII}(a) (high $V$).
On the other hand, from Fig.~\ref{fig:resultsG2LO} we found that in this limit photons at the Josephson frequency appear also almost randomly,
with a slight antibunching.


\begin{figure}[tb]
\includegraphics[width=\linewidth]{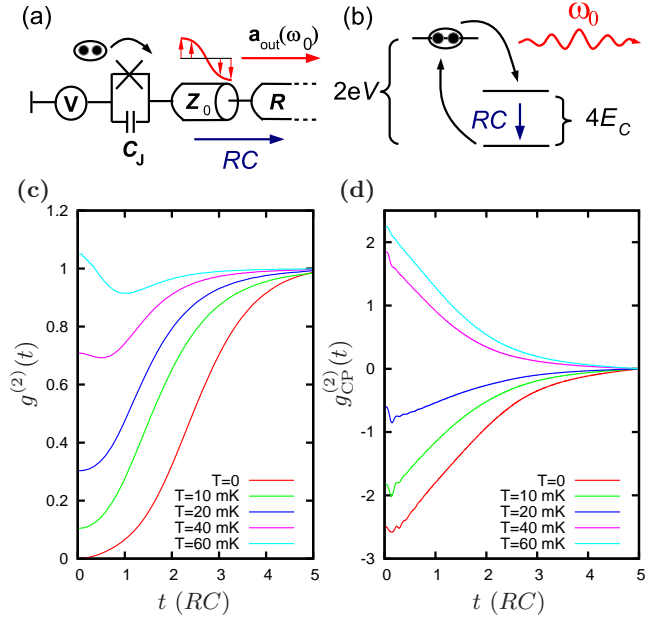}
\caption{
(a) We consider a transmission line with a stepped characteristic impedance providing a high zero-frequency impedance ($R=4R_{\rm Q}$) and a resonance at $\omega_0/2\pi=5$~GHz.
(b) When biased at $2eV=4e^2/2C+\hbar\omega_0$, the charge transport is expected to occur through repeated single-photon assisted Cooper-pair tunneling
with a finite $RC$ charging time in between.
(c) Second-order coherence of the photons emitted at the resonance frequency $\omega_0$~\cite{Leppakangas20152}.
At temperatures roughly below $3E_C/k_{\rm B}\approx 60$~mK we observe antibunching within the $RC$-timescale.
We measure photons at the resonance frequency within bandwidth $\Delta/\pi=1$~GHz.
(d) Second-order coherence of Cooper-pair transport as defined via low-frequency current noise, Eq.~(\ref{eq:SecondCoherenceVoltage}).
The $g^{(2)}_{\rm CP}(t)$-function describes simultaneous changes in the low-frequency emission due to
correlations between consecutively tunneling Cooper pairs.
}
\label{fig:resultsG2HO}
\end{figure}

We continue by investigating the second-order coherences in the Coulomb blockade regime. 
For this purpose we consider a stepped characteristic impedance,
which provides a $\lambda/2$ standing-wave between the junction and the step, see Fig.~\ref{fig:resultsG2HO}(a)
(modifications to main formulas are considered in Appendix A).
The setup is equivalent to a driven cavity with a small leakage and essentially
increases radiation  in the transmission line nearby the resonance frequency $\omega_0$.
The photon antibunching in this setup has been studied earlier in Ref.~[\onlinecite{Leppakangas20152}].
In Fig.~\ref{fig:resultsG2HO}(c), we show the obtained results for the second-order coherence of photons at $\omega_0$.
We consider a voltage bias that provides resonant single-photon assisted Cooper-pair tunneling, $2eV=4e^2/2C+\hbar\omega_0$.
The capacitance $C=50$~fF is here defined by the low-frequency cut-off frequency and is approximately double the real junction capacitance $C_{\rm J}$~\cite{Leppakangas20152}.
The main result is that we observe antibunching approximately  when the temperature is smaller than
three times the charging energy, $k_{\rm B}T < 3E_C$.


In  Fig.~\ref{fig:resultsG2HO}(d), we plot the corresponding second-order coherence of Cooper-pair transport, $g^{(2)}_{\rm CP}(t)$, defined in Eq.~(\ref{eq:SecondCoherenceVoltage}).
This describes changes in the low-frequency current-noise due to correlations between consecutively tunneling Cooper pairs.
For simplicity, we filter the current fluctuations through a low-pass $RC$-filter instead of $\vert A(\omega)\vert^2$.
We see that at low temperatures the photon antibunching indeed occurs
together with a decrease in the zero-frequency noise, $\int dt g^{(2)}_{\rm CP}(t)<0$.
The resulting $g^{(2)}_{\rm CP}(t)$-function is also characterized by an exponential-type recovery.
The relevant timescale here is again the charging time $RC$.
This is consistent with the idea of Coulomb blockade of Cooper-pair tunneling during the junction recharging.
An analogous effect was found in the case of a bare high-Ohmic environment, see Section~\ref{sec:G2CP}.
At short times the $g^{(2)}_{\rm CP}(t)$-function is superposed by high-frequency oscillations.
These are a remnant of the filtered high-frequency fluctuations,
describing changes in the emission spectrum nearby the mode frequency $\omega_0$.
When increasing the temperature
a "bunching" effect  emerges, $\int dt g^{(2)}_{\rm CP}(t)>0$, similarly as in the case of photon statistics.
However, this occurs well before the bunching of the photons at $\omega_0$.
This is understood within semiclassical analysis as a result of an
increased probability for consecutive Cooper-pair tunneling without emission of a second photon.


\section{Conclusions and Outlook}\label{sec:discussion}
In this article we developed theoretical tools for the study of
continuous-mode microwave radiation emitted by inelastic Cooper-pair tunneling.
We considered explicitly a voltage-biased Josephson junction terminating a semi-infinite transmission line.
The evaluation of the system properties up to the fourth-order in the tunneling coupling
accounted for nonequilibrium dynamics between consecutively tunneling Cooper pairs.
Particularly, we investigated how the interaction between consecutively tunneling Cooper pairs
affects the finite-frequency current-noise and the second-order coherence of the emitted photons.
We also addressed the general connection between the junction current noise, voltage noise, radiation power,
and the first-order coherence of the emitted photons.
The fourth-order approach accesses phenomena emerging in weak non-equilibrium in the dynamical Coulomb blockade.

This formalism is a complementary tool for investigating nonclassical microwave production in Josephson-junction-resonator systems, for example,
among the versatile Jaynes-Cummings type approaches~\cite{Marthaler2011,Ulm2013,Nottingham2013}.
It is particularly useful for investigating the regime of weak photon emission,
as it is able to model arbitrary forms of the electromagnetic environment, including the "flat" Ohmic transmission line.
It also directly addresses an important experimentally accessible observable, the continuous-mode emission field,
and naturally accounts for thermal noise and dephasing present in all experiments.
The regime of weak photon flux is very relevant in certain quantum-information applications,
including production of single-photons and entangled photon pairs.

Finally, we found close relations between different moments of the junction current and the second-order coherence of the emitted radiation.
This raises motivation for further studies of such relations and their form in situations beyond the perturbative regime.
The question asked
is whether microwave correlation measurements can also give new information about counting statistics and related properties of the mesoscopic charge transport.

\section*{Acknowledgments}
We thank M.~Hofheinz, F.~Portier, A.~Grimm, and O. Parlavecchio for numerous motivating discussions.
This work was financially supported by the Swedish Research Council and by the EU STREP project PROMISCE.


\appendix*

\section*{Appendix A: General transmission line}
\begin{figure}[tb]
\includegraphics[width=\linewidth]{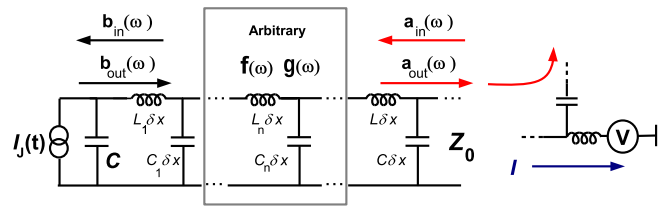}
\caption{
In the more general analysis we consider an arbitrary transmission line section that is
from one side terminated by the Josephson junction and from the other side
connected to a semi-infinite transmission line with a constant characteristic impedance $Z_0$.
In an experimental setup the outgoing high-frequency radiation can be
measured independently from the direct current by using a bias-T~\cite{Hofheinz2011}.
}
\label{fig:Generalization}
\end{figure}

Here we study changes to main formulas when the nearby electromagnetic environment has resonance frequencies.
The explicit setup we consider is shown in Fig.~\ref{fig:Generalization}.
At certain distance from the Josephson junction the emitted radiation leaks to a semi-infinite transmission line ($Z_0$),
wherefrom its properties can be measured.
At the chip the nearby microwave circuit is otherwise arbitrary but ends to a segment of transmission line
with a characteristic impedance $Z_1$, which is then terminated by the junction.
Applying classical electric-circuit analysis,
the region between the free space and the junction can be described by transfer
functions $g(\omega)$ and $f(\omega)$, which connect the incoming and outgoing components of the traveling wave.
The same functions also the connect the corresponding field operators and we have
\begin{eqnarray}
\hat b_{\rm out}(\omega)  &=& f(\omega) \hat a_{\rm in}(\omega) + g(\omega) \hat a_{\rm out}(\omega)  \label{CavityBoundary1} \\
\hat b_{\rm in}(\omega)  &=& g^*(\omega)\hat a_{\rm in}(\omega)  + f^*(\omega) \hat a_{\rm out}(\omega) \label{CavityBoundary2}.
\end{eqnarray}
Here  $\hat b(\omega)$ refers to the field operator beside the junction, see Fig.~\ref{fig:Generalization}.
The functions $g(\omega)$ and $f(\omega)$  depend on the realization of the circuit,
but always satisfy the condition $\vert g(\omega)\vert^2 -\vert f(\omega)\vert^2=1$.
This relation is valid for a lossless and reciprocal network~\cite{Pozar}.

To evaluate the junction phase fluctuations for $I_{\rm J}=0$,
we Fourier transform the boundary condition at the junction and obtain
\begin{eqnarray}
c(\omega)\hat b_{\rm out}(\omega) - c^*(\omega)\hat b_{\rm in}(\omega)=0,
\end{eqnarray}
where we have defined $c(\omega)=1-iZ_1C_{\rm J}\omega $.
The solution is
$\hat b_{\rm out}(\omega)= e^{i\theta}\hat b_{\rm in}(\omega)$,
where $e^{i\theta}=c^* (\omega)/c(\omega)$.
Combining this with equations~(\ref{CavityBoundary1}-\ref{CavityBoundary2}) we obtain
\begin{eqnarray}
\hat a_{\rm out}&=&\frac{f(\omega)e^{-i\theta}-g^*(\omega)}{f^*(\omega)-g(\omega)e^{-i\theta}}\hat a_{\rm in}.
\end{eqnarray}
The operator describing the phase difference across the Josephson junction is now propotional to the sum
\begin{eqnarray}
\hat b_{\rm in}+\hat b_{\rm out}&=&\hat b_{\rm in}(1+e^{i\theta})\\ \nonumber
&=& (1+e^{i\theta})\left(  \frac{\vert f(\omega)\vert^2 e^{-i\theta}-\vert g(\omega)\vert^2e^{-i\theta} }{f^*(\omega)-g(\omega)e^{-i\theta}} \right) \hat a_{\rm in}(\omega)\\ \nonumber
&=& \frac{2}{c(\omega)g(\omega) -c^*(\omega) f^*(\omega) } \hat a_{\rm in}(\omega).
\end{eqnarray}
The factor in front of the operator $\hat a_{\rm in}(\omega)$ is proportional to the modified form of $A(\omega)$.
To check the consistency of this we also solve the out-field $\hat a_{\rm out}(\omega)$ resulting from a tunneling current.

In the region of $Z_0$ we have no input field from the semi-infinite transmission line that would originate from the tunneling current.
Therefore, for finite orders in $E_{\rm J}$ Eqs.~(\ref{CavityBoundary1}-\ref{CavityBoundary2}) get the form
\begin{eqnarray}
\hat b_{\rm out}(\omega)&=&g(\omega)\hat a_{\rm out}(\omega), \\
\hat b_{\rm in}(\omega)&=&f^*(\omega)\hat a_{\rm out}(\omega) .
\end{eqnarray}
The field at the junction has to satisfy,
\begin{eqnarray}
\hat b_{\rm out}(\omega)=\hat b_{\rm in}(\omega)e^{i\theta(\omega)}+ i \frac{ \sqrt{ Z_1/\hbar\omega\pi } }{c(\omega)  }\int_{-\infty}^{\infty}e^{i\omega t}dt \hat I_{\rm J}(t). \nonumber
\end{eqnarray}
It follows then (for finite orders in $E_{\rm J}$)
\begin{eqnarray}
&&\hat a_{\rm out}(\omega)=\\
&&i \sqrt{\frac{Z_1}{\hbar\omega\pi}} \frac{1}{[c(\omega)g(\omega)-c^*(\omega)f^*(\omega)]}\int_{-\infty}^{\infty}e^{i\omega t}dt \hat I_{\rm J}(t). \nonumber
\end{eqnarray}

From the two above calculations we deduce that the phase fluctuations and the output field are both described as in the main part of the article,
but with the redefinition
\begin{equation}
A(\omega)= \sqrt{\frac{Z_1}{Z_0}}\frac{1}{[c(\omega)g(\omega)-c^*(\omega)f^*(\omega)]}.
\end{equation}
The discussed Ohmic transmission line is obtained as the special case $g(\omega)=1$, $f(\omega)=0$, and $Z_1=Z_0$.
The step impedance considered in Section~\ref{sec:NumericalPhotons} consists of impedance $Z_1$ of length $d$ connected to the open line with $Z_0$.
This is described by $g=(\sqrt{Z_1/Z_0}+\sqrt{Z_0/Z_1})/2$ and $f=(\sqrt{Z_0/Z_1}-\sqrt{Z_1/Z_0})e^{-2ik_\omega^1 d}/2$.

\section*{Appendix B: Derivation of $G^{(2)}(\tau)$}
Here we show how
the different terms and the Keldysh time-ordering emerge in the derivation of the term $G^{(2)}$,
defining the second-order coherence.
We take use of the solution for $\hat a_0(\omega)$,
\begin{equation}\label{ZerothOrderField}
\hat a_0(\omega)=\frac{\Phi_0}{2\pi}\frac{1}{ A^*(\omega)}\sqrt{\frac{\omega}{4\pi\hbar Z_0}}\int_{-\infty}^{\infty}e^{i\omega t}\hat \phi_0(t).
\end{equation}
We also use
\begin{eqnarray}\label{MixedTerms}
&&\left\langle e^{-i\hat\phi}\hat\phi'\hat \phi''e^{i \hat \phi'''}  \right\rangle=\left\langle e^{-i\hat\phi}e^{i\hat\phi'''}  \right\rangle \times \\
&& \left\{ \left\langle \hat\phi'\hat\phi'' \right\rangle +\left[  \left\langle \hat\phi''\hat\phi'''\right\rangle-\left\langle \hat\phi\hat\phi'' \right\rangle  \right] \left[  \left\langle \hat\phi\hat\phi' \right\rangle  -\left\langle \hat\phi'\hat\phi''' \right\rangle \right]   \right\}.\nonumber
\end{eqnarray}
Here we have simplified the notation by using $\hat \phi\equiv \hat\phi_0(t)$.
We note that the bare operators $\hat\phi'$ and  $\hat \phi''$ become contracted both to  $\hat\phi$ (left) and to $ \hat \phi'''$ (right).
It turns out that the free evolution $\phi'$ has to be paired to left, and $\phi''$ to right in order to have contribution
at zero temperature. The relevant in the following is then
\begin{eqnarray}\label{MixedTerms}
\lim_{T\rightarrow 0} \left\langle e^{-i\hat\phi}\hat\phi'\hat \phi''e^{i \hat \phi'''}  \right\rangle =\left\langle e^{-i\hat\phi}e^{i\hat\phi'''}  \right\rangle \  \left\langle \hat\phi\hat\phi' \right\rangle\ \left\langle \hat\phi''\hat\phi'''\right\rangle  . \nonumber
\end{eqnarray}

We can generalize this to higher order terms by
noting that $e^{i\hat\phi(t)}e^{i\hat\phi(x)}=e^{i\hat\phi(t)+i\hat\phi(x)}c(t-x)$,
where $c(t)$ is a complex number, which can be used to deduce that
\begin{eqnarray}\label{MixedTerms}
&&\lim_{T\rightarrow 0}\left\langle e^{-i\hat\phi} e^{-i\hat\phi^x} \hat\phi'\hat \phi'' e^{i\hat\phi^y} e^{i \hat \phi'''}  \right\rangle\\
&&= \left\langle e^{-i\hat\phi} e^{-i\hat\phi^x} e^{i\hat\phi^y}e^{i\hat\phi'''}  \right\rangle   \left\langle ( \hat\phi+ \hat\phi^x) \hat\phi' \right\rangle  \left\langle \hat\phi''  (\hat\phi^y+ \hat\phi''') \right\rangle  . \nonumber
\end{eqnarray}
In the next step we use the result~(\ref{eq:PhaseCorrelations}) for the phase correlations,
which leads to
\begin{eqnarray}
\int_{-\infty}^{\infty}e^{-i\omega' t'} \left\langle\hat \phi_0(x)\hat \phi_0(t')\right\rangle=\frac{4\pi}{\omega'}\frac{  Z_0\vert A(\omega')\vert^2 }{R_Q}\frac{e^{-i\omega'x}}{1-e^{-\beta\hbar\omega'}}. \nonumber
\end{eqnarray}
We get for $\left\langle \hat a_2^{\dagger}(\omega)\hat a_0^{\dagger}(\omega')\hat a_0(\omega'')\hat a_2(\omega''') \right\rangle$
(that we name from now on as the term [2002], similarly for others)
\begin{eqnarray}\label{eq:2002}
&&\int_{-\infty}^{\infty} dt \int_{-\infty}^{\infty} dt'''\int_{-\infty}^{t}dx \int_{-\infty}^{t'''}dy \\
&& \times \left\langle [e^{-i\hat\phi},e^{-i\hat\phi^x}][e^{i\hat\phi^y},e^{i\hat\phi'''}] \right\rangle \nonumber\\
&&\times \frac{I_{\rm c}^4Z_0^2 A^*(\omega)A^*(\omega') A(\omega'')A(\omega''')}{\pi^2\hbar^2\sqrt{\omega\omega'\omega''\omega'''}}\nonumber\\
&& \times  e^{-it\omega}e^{it'''\omega'''} \left( e^{-i\omega' x} + e^{-i\omega' t}\right)\left( e^{+i\omega'' y} + e^{+i\omega'' t'''}\right). \nonumber
\end{eqnarray}
The front factor is the same as in the term [1111],  $\left\langle \hat a_1^{\dagger}(\omega)\hat a_1^{\dagger}(\omega')\hat a_1(\omega'')\hat a_1(\omega''') \right\rangle$. However, in [2002] a time-ordering appears and the possibility
to have various energy factors (different terms after  the expansion of the two multiplied parentheses).
From [2002] we pick up a term similar to [1111] [other terms contribute to ${\cal I}_2(\tau)$ and ${\cal I}_3(\tau)$]
\begin{eqnarray}\label{eq:term11}
&&\left\langle \hat a_2^{\dagger}(\omega)\hat a_0^{\dagger}(\omega')\hat a_0(\omega'')\hat a_2(\omega''')\right\rangle \rightarrow \\
&&+{\cal T}(t' \rightarrow t) \left\langle \hat a_1^{\dagger}(\omega)\hat a_1^{\dagger}(\omega')\hat a_1(\omega'')\hat a_1(\omega''') \right\rangle {\cal T}(t'' \rightarrow t''')\label{eq:term1} \nonumber \\
&&+ {\cal T}(t' \rightarrow t) \left\langle\hat a_1^{\dagger}(\omega')\hat a_1^{\dagger}(\omega)\hat a_1(\omega''')\hat a_1(\omega'') \right\rangle {\cal T}(t'' \rightarrow t''')\label{eq:term2} \nonumber  \\
&&-{\cal T}(t' \rightarrow t) \left\langle \hat a_1^{\dagger}(\omega)\hat a_1^{\dagger}(\omega')\hat a_1(\omega''')\hat a_1(\omega'') \right\rangle {\cal T}(t'' \rightarrow t''')\label{eq:term3} \nonumber \\
&&- {\cal T}(t' \rightarrow t) \left\langle\hat a_1^{\dagger}(\omega')\hat a_1^{\dagger}(\omega)\hat a_1(\omega'')\hat a_1(\omega''') \right\rangle {\cal T}(t'' \rightarrow t'''). \label{eq:term4} \nonumber
\end{eqnarray}
Here ${\cal T}$ stands for time ordering. The time-ordering is always the same
but the terms inside the expectation value exchange their positions.
The first term on the right-hand side has the same energy-armuments as the term [1111], but has a fixed time ordering.

Similar calculation can also be done for terms [1102] and [2011]. We use here the relation
\begin{eqnarray}\label{MixedTerms}
\left\langle e^{-i\hat\phi}\hat \phi''e^{i \hat \phi'''}  \right\rangle=\left\langle e^{-i\hat\phi}e^{i\hat\phi'''}  \right\rangle \ \left[ -i\left\langle \hat\phi\hat\phi'' \right\rangle\ +i \left\langle \hat\phi''\hat\phi'''\right\rangle \right] . \nonumber
\end{eqnarray}
We note that
when contracting to left, a minus sign appears.  The zero-temperature limit gives
\begin{eqnarray}\label{eq:term22}
&&\left\langle\hat a_1^{\dagger}(\omega)\hat a_1^{\dagger}(\omega')\hat a_0(\omega'')\hat a_2(\omega''')\right\rangle \\
&&+ \left\langle\hat a_2^{\dagger}(\omega)\hat a_0^{\dagger}(\omega')\hat a_1(\omega'')\hat a_1(\omega''')\right\rangle \nonumber \\
&&=- \left\langle\hat a_1^{\dagger}(\omega)\hat a_1^{\dagger}(\omega')\hat a_1(\omega'')\hat a_1(\omega''') \right\rangle {\cal T}(t'' \rightarrow t''')\label{eq:term5} \nonumber \\
&&-{\cal T}(t' \rightarrow t) \left\langle\hat a_1^{\dagger}(\omega)\hat a_1^{\dagger}(\omega')\hat a_1(\omega'')\hat a_1(\omega''') \right\rangle \label{eq:term6}\nonumber \\
&&+ \left\langle\hat a_1^{\dagger}(\omega)\hat a_1^{\dagger}(\omega')\hat a_1(\omega''')\hat a_1(\omega'') \right\rangle {\cal T}(t'' \rightarrow t''') \label{eq:term7} \nonumber\\
&&+ {\cal T}(t' \rightarrow t) \left\langle\hat a_1^{\dagger}(\omega')\hat a_1^{\dagger}(\omega)\hat a_1(\omega'')\hat a_1(\omega''') \right\rangle.\label{eq:term8} \nonumber
\end{eqnarray}
We obtain that the first and the second term 
kill all time-orderings from [1111], except the natural ones where inner operors follow timewise
the outer operators ($t'>t$ and $t''>t'''$). The second term 
on the right-hand side of Eq.~(\ref{eq:term22}) on the other hand gives the same term but with interchanged energy-arguments
($\omega\rightarrow\omega'$ {\em and} $\omega''\rightarrow\omega'''$). The rest of the terms give the same contribution
for ($\omega\rightarrow\omega'$ {\em or} $\omega''\rightarrow\omega'''$).
After summation of all possible terms, the overall integration can be rewritten in the compact form
\begin{eqnarray}
&& \int_{-\infty}^{\infty}dt \int_{-\infty}^{\infty}dt' \int_{-\infty}^{\infty}dt'' \int_{-\infty}^{\infty}dt''' \nonumber \\
&&\times \left \langle {\cal T}^{\dagger}\left\{ \hat I^{\dagger}_{\omega,t} \hat I^{\dagger}_{\omega',t'} \right\}  \  {\cal T}\left\{\hat I_{\omega'',t''} \hat I_{\omega''',t'''}\right\} \right \rangle  \nonumber.
\end{eqnarray}
We find that the Keldysh time-ordering appears in the final expression only when summation over all terms in the fourth order is performed.

\section*{Appendix C: Evaluation of ${\cal I}_4(\tau)$}
At the center of the explicit evaluation of ${\cal I}_4(\tau)$ is the  correlator
\begin{eqnarray}\label{Eq:BunchingCorrelator}
&&  \left\langle e^{-i\phi}   e^{-i\phi'}  e^{i\phi''} e^{i\phi'''}  \right\rangle =\\
&&\exp\left[ J(t-t'')+J(t-t''')+J(t'-t'')+ \right. \nonumber\\
&& \left. J(t'-t''')-J(t-t')-J(t''-t''') \right] \nonumber .
\end{eqnarray}
We include here only forward directed Cooper-pair tunnellings, meaning $\hat a\propto  e^{i\hat\phi}$, other directions are treated similarly.
We need to perform time integration over the expression~(\ref{Eq:BunchingCorrelator})
multiplied by specific energy factors given below.
To do this we define a new set of variables,
\begin{eqnarray}
x&=&t'-t'''   \,\,\,\,\,\,\,\,\,\,\,\,\,\,\,\,\,\,\,\,\,\,\,\,\,\,\,\,\,\,\,  t'=\frac{2x+y-z+4s}{4} \nonumber  \\
y&=&t'-t''  \,\,\,\,\,\,\,\,\,\,\,\,\,\,\,\,\,\,\,\,\,\,\,\,\,\,\,\,\,\,\,\,\,  t=\frac{-2x+y+3z+4s}{4} \nonumber \\
z&=&t-t''' \,\,\,\,\,\,\,\,\,\,\,\,\,\,\,\,\,\,\,\,\,\,\,\,\,\,\,\,\,\,\,  t''=\frac{2x-3y-z+4s}{4}  \nonumber \\
s&=&\frac{t+t'+t''+t'''}{4} \,\,\,\,\,  t'''=\frac{-2x+y-z+4s}{4} \nonumber
\end{eqnarray}
We get for the correlator on the right-hand side of equation~(\ref{Eq:BunchingCorrelator}),
\begin{eqnarray}\label{eq:4thOrderCorrelator2}
e^{J(y)+J(z)}\left[ e^{J(x)+J(y+z-x)-J(x-y)-J(z-x)}\right].
\end{eqnarray}
This has no dependence on $s$ and
when integrating over $s$ we get a term $2\pi \delta(\omega+\omega'-\omega''-\omega''')$, accounting for energy conservation.

A central role in the following is played by the term inside the large parentheses of Eq.~(\ref{eq:4thOrderCorrelator2}).
The asymptotic behaviour
of this term for $x\rightarrow \infty$ is 1.
This means that integration over $x$ seems to produce an extra $\delta$-function.
A more careful analysis shows that this holds when $x>y,z$, or $x<y,z$,
the integration over $x$ should be done with this restriction.
This means $t'>t$ and $t''>t'''$ (Keldysh time-ordering), or the opposite. 

We proceed to perform integration over $x$. We need to multiply the correlator by an additional energy-factor of type,
\begin{eqnarray}
E(x,y,z)&=&e^{ix(\omega''-\omega')}\, e^{iy(\omega_{\rm J}-\omega'')}\, e^{iz(\omega_{\rm J}-\omega)}.
\end{eqnarray}
We have used here the condition coming from the $s$-integration, $\omega+\omega'=\omega''+\omega'''$.
It turns out to be numerically convenient to consider the sum of the two integration orders,
$x>y,z$ and $x<y,z$ (not only the first one).
This can be mapped to complex conjugation of the integral. In the following we take use of the property
\begin{eqnarray}\label{eq:EvaluationOfI4}
&&\left[A_\omega\int_{{\rm Max}(y,z)}^{\infty}dx + A_\omega^*\int^{{\rm Min}(y,z)}_{-\infty}dx \right]  e^{x(\omega''-\omega')} \\
&& =  2\pi \vert A_\omega\vert \delta (\omega''-\omega') +\nonumber \\
&&+  i \left(  A_\omega e^{i(\omega''-\omega'){\rm Max}(y,z)} - A_\omega^* e^{i(\omega''-\omega'){\rm Min}(y,z)} \right) \frac{\cal P}{\omega''-\omega'}.\nonumber
\end{eqnarray}
Here $\cal P$ stands for principal value integration.
The $\delta$-function can now be integrated over $y$ and $z$,
leading to a contribution of type
\begin{equation}\label{eq:newRef1}
\frac{1}{2} s(\omega,\omega''')s(\omega',\omega''),
\end{equation}
where $s(\omega,\omega')$ is the leading-order result for the emission-power density, Eq.~(\ref{eq:PowerDensityNew}).
When exchanging $\omega \leftrightarrow\omega'$ and $\omega'' \leftrightarrow\omega'''$, that emerges due to the time-ordering in Eq.~(\ref{eq:G2MainContribution}),
we get the same result, which ultimately removes the factor $1/2$. When exchanging only one of the energy arguments we get two contributions of type
\begin{equation}\label{eq:newRef2}
\frac{1}{2} s(\omega,\omega'')s(\omega',\omega''').
\end{equation}
These are the dominating contribution in the zero bandwidth limit, ${\rm BW}\rightarrow 0$.

Finite-bandwidth correlations are described by terms beyond Eqs.~(\ref{eq:newRef1}-\ref{eq:newRef2}).
In the main part of the article we marked their contribution as the function ${\cal C}(\omega_1,\omega_2,\omega_3,\omega_4)$,
whose analytical form is now
\begin{eqnarray}
&&{\cal C}=  \left( \frac{Z_0I_{\rm c}^2}{4\pi} \right)^2 \times \\
&&\sum_{\omega_1\leftrightarrow\omega_2,\omega_3\leftrightarrow\omega_4}C(\omega_1,\omega_2,\omega_3,\omega_4)\delta(\omega_1+\omega_2-\omega_3-\omega_4) \nonumber \\
&&C=i \int_{-\infty}^{\infty}dy\int_{-\infty}^{\infty}dz e^{J(y)}e^{J(z)} \\
&&\times \left( A_\omega e^{i(\omega''-\omega'){\rm Max}(y,z)} -A_\omega^* e^{i(\omega''-\omega'){\rm Min}(y,z)} \right)\nonumber\\
&&\times \frac{\cal P}{\omega''-\omega'}e^{i(\omega_{\rm J}-\omega'')y}e^{i(\omega_{\rm J}-\omega)z}\nonumber\\
&&+\int_{-\infty}^{\infty} dy\int_{-\infty}^{\infty} dz F(y,z) e^{J(y)} e^{J(z)} e^{iy(\omega_{\rm J}-\omega'')}e^{iz(\omega_{\rm J}-\omega)}. \nonumber
\end{eqnarray}
Here the summation is performed over exchanging the energy arguments as indicated (in total four terms).
We have also introduced a term
\begin{eqnarray}\nonumber
&&F(y,z)\equiv \left[A_\omega\int_{{\rm Max}(y,z)}^{\infty}dx + A_\omega^*\int^{{\rm Min}(y,z)}_{-\infty}dx \right] \nonumber\\
&&\times \left[e^{J(x)+J(y+z-x)-J(x-y)-J(z-x)} - 1 \right]e^{i(\omega''-\omega')x}. \nonumber
\end{eqnarray}
The results presented in Section~\ref{sec:NumericalPhotons} are based on numerical evaluation of $C$
using two- and three-dimensional fast Fourier transformations.

\begin{figure}[bt]
\includegraphics[width=0.65\linewidth]{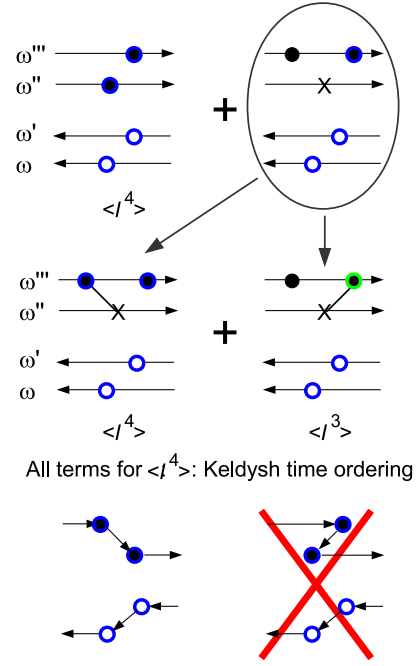}
\caption{Visualization of full  calculation of  $G^{(2)}$. Each contribution in the summation~(\ref{eq:GeneralCorrelator})
is given a four-branched diagram. The two upper branches correspond to photon annihilation
operators, $\hat a(\omega_3)$ and $ \hat a(\omega_4)$, and the two lower to creation operators,
$\hat a^{\dagger}(\omega_1)$ and $ \hat a^{\dagger}(\omega_2)$.
The circles describe timings of tunneling events in operators $\hat a_n$ ($n\geq 1$)
(see  Section~\ref{sec:KeldyshCalculation}) and letters X are related to the vacuum noise ($n=0$).
Initially, only the last tunneling of $\hat a_n$ ($n\geq 1$) is associated with Fourier frequency, marked as the blue shell,
whereas circles with no extra shell describe previous (time-ordered) evolution on the same branch.
There is no time-ordering between points at different branches.
In the ensemble average, the vacuum-noise operators (X) contribute only when placed to inner branches.
There they become connected to the circles of the same time direction, changing them either to blue (with Fourier frequency
of the inner branch), or to green (with the sum of Fourier frequencies of the two brances). The total number of
coloured shells gives the order of the current moment this expression is related.
After summation over all terms a Keldysh time-ordering appears, which allows expressing the result using only two branches.}
\label{fig:KeldyshPhotonBunching}
\end{figure}

\section*{Appendix D: Terms ${\cal I}_2(\tau)$ and ${\cal I}_3(\tau)$}
The full calculation of $G^{(2)}(\tau)$ is visualized in Fig.~\ref{fig:KeldyshPhotonBunching}.
We see that also other contributions than ${\cal I}_4(\tau)$ appear when disentangling the phase operators in all possible ways.
Particularly, a set of contributions [which we mark ${\cal I}_2(\tau)$] can be mapped to the second moment of the junction current.
In Fig.~\ref{fig:KeldyshPhotonBunching},
this corresponds to terms where the inner operators $\hat a^\dagger(\omega_2)$ and $\hat a(\omega_3)$ are of zeroth order and
connected to the
green points on the outer branches. The green points correspond to picking the energy factors
$e^{-i\omega' t}$ and $e^{i\omega''t'''}$ in contribution~(\ref{eq:2002}).
This type of a relation turns out to hold in all orders. We get
\begin{eqnarray}\label{eq:G2II}
 {\cal I}_2(\tau) &=&  \left(\frac{\hbar Z_0}{4 \pi}\right)^2 \frac{2}{\pi\hbar R_{\rm Q}} \int_{\rm BW} A_\omega \int_{\rm times} e^{i\tau(\omega'-\omega'')}\nonumber  \\
&\times& \left\langle \hat I_{\rm J}(t_1) \hat I_{\rm J}(t_4)\right\rangle e^{it_1(\omega_1+\omega_2)}e^{-it_4(\omega_3+\omega_4)} . \nonumber
\end{eqnarray}
This expression can be further recasted in the form
\begin{eqnarray}\label{eq:G2II2}
{\cal I}_2(\tau) &=& \left(\frac{\hbar Z_0}{4 \pi}\right)^2 \frac{4}{\hbar R_{\rm Q}} \int_{\rm BW} A_\omega e^{i\tau(\omega'-\omega'')} \\
&\times& \left\langle \hat I_{\rm J}(t) \hat I_{\rm J}(0)\right\rangle_{\omega_1+\omega_2} \delta(\omega_1+\omega_2-\omega_3-\omega_4) .  \nonumber
\end{eqnarray}
In Section~\ref{sec:power}, this type of an expression was  evaluated up to the fourth-order in the critical current.

Similarly, a set of contributions [which we mark ${\cal I}_3(\tau)$] can be mapped to evaluation of third moment of the current fluctuations
when expanded to fourth order in the tunneling coupling.
A contribution here can be written in the qualitative form
\begin{eqnarray}
&&\int_{\rm times}{\cal T}^{\dagger}\left\{ \hat I_{\omega,t}\ \hat I_{\omega',t'} \right\}\ \hat I_{\rm J}(t''') e^{-it'''(\omega''+\omega''')} +  \nonumber\\
&&\ \int_{\rm times} \hat I_{\rm J}(t)e^{it(\omega+\omega')}\ {\cal T}\left\{ \hat I_{\omega'',t''}\ \hat I_{\omega''',t'''} \right\}.
\end{eqnarray}
Here the operators $\hat I(t)$ and $\hat I(t''')$ are expanded to the second order in the tunnel coupling.
The other terms have also a similar form
\begin{eqnarray}
&&\int_{\rm times}{\cal T}^{\dagger}\left\{ \hat I_{\omega,t}\ \hat I_{\omega',t'} \right\}\hat I_{t''',\omega''+\omega'''} \nonumber\\
&& + \ \hat I_{t,\omega+\omega'}  \int_{\rm times}{\cal T}\left\{ \hat I_{\omega'',t''}\ \hat I_{\omega''',t'''} \right\}.
\end{eqnarray}
Here the operators inside the braces are expanded in the third order of the tunneling coupling.
The expansion is made by inserting a free evolution term
before, between, or after the
time-ordered operators $\hat I_{\omega,t}$.
A more detailed analysis of these type of contributions is a subject of future works.



\begin{thebibliography}{99}

\bibitem{Devoret1990}
M.~H.~Devoret, D.~Esteve, H.~Grabert, G.-L.~Ingold, H.~Pothier, and C.~Urbina, Phys.~Rev.~Lett.~{\bf 64}, 1824 (1990).


\bibitem{Girvin1990}
S. M. Girvin, L. I. Glazman, M. Jonson, D. R. Penn, and M. D. Stiles, Phys.~Rev.~Lett.~{\bf 64}, 3183 (1990).

\bibitem{Ingold1992}
G.-L.~Ingold and Yu.~V.~Nazarov, in {\em Single Charge Tunneling: Coulomb Blockade Phenomena in Nanostructures}, edited by H.~Grabert and M.~H.~Devoret (Plenum, New York, 1992), p.21.

\bibitem{Holst1994}
T.~Holst, D.~Esteve, C.~Urbina, and M.~H.~Devoret, Phys.~Rev.~Lett.~{\bf 73}, 3455 (1994).


\bibitem{Hofheinz2011}
M.~Hofheinz, F.~Portier, Q.~Baudouin, P.~Joyez, D.~Vion, P.~Bertet, P.~Roche, and D.~Esteve, Phys.~Rev.~Lett.~{\bf 106}, 217005 (2011).


\bibitem{Leppakangas2013}
J.~Lepp\"akangas, G.~Johansson, M.~Marthaler, and M. Fogelstr\"om, Phys.~Rev.~Lett.~{\bf 110}, 267004 (2013).

\bibitem{Nottingham2013}
A.~D.~Armour, M.~P.~Blencowe, E.~Bahimi, A.~J.~Rimberg, Phys. Rev. Lett.~{\bf 111} 247001 (2013).

\bibitem{Ulm2013}
V.~Gramich, B.~Kubala, S.~Rocher, J.~Ankerhold, Phys. Rev. Lett.~{\bf 111} 247002 (2013).

\bibitem{Ulm2015}
A. D. Armour, B. Kubala, and J. Ankerhold,  Phys. Rev. B {\bf 91}, 184508 (2015).

\bibitem{Paris2015}
M. Trif and P. Simon, Phys. Rev. B {\bf 92}, 014503 (2015).

\bibitem{Leppakangas20152}
J.~Lepp\"akangas, M. Fogelstr\"om, A.~Grimm, M.~Hofheinz, M.~Marthaler, and G.~Johansson, Phys. Rev. Lett.~{\bf 115} 027004 (2015).

\bibitem{Dambach2015}
S. Dambach, B. Kubala, V. Gramich, and J. Ankerhold,  Phys. Rev. B {\bf 92}, 054508 (2015).


\bibitem{Beenakker2001}
C. W. J. Beenakker and H. Schomerus,  Phys. Rev. Lett.~{\bf 86} 700 (2001).

\bibitem{Beenakker2004}
C. W. J. Beenakker and H. Schomerus,  Phys. Rev. Lett.~{\bf 93} 096801 (2004).


\bibitem{Portier2}  
O. Parlavecchio, C. Altimiras, J.-R. Souquet, P. Simon, I. Safi, P. Joyez, D. Vion, P. Roche, D. Esteve, and F. Portier,
Phys. Rev. Lett.~{\bf 114} 126801 (2015).


\bibitem{Jin2011}
P.-Q. Jin, M. Marthaler, J. H. Cole, A. Shnirman, and G. Sch\"on, Phys. Rev. B {\bf 84}, 035322 (2011).

\bibitem{Nazarov2012}
C.~Padurariu, F.~Hassler, and Y.~V.~Nazarov, Phys.~Rev.~B~{\bf 86}, 054514 (2012).

\bibitem{Marthaler2011}
M. Marthaler, J. Lepp\"akangas, and J. H. Cole, Phys. Rev. B {\bf 83}, 180505(R) (2011).

\bibitem{Leppakangas2014}
J.~Lepp\"akangas, G.~Johansson, M.~Marthaler, and M. Fogelstr\"om, New J.~Phys.~{\bf 16}, 015015 (2014).

\bibitem{NatureCommunications2014}
J.-R. Souquet, M.J. Woolley, J. Gabelli, P. Simon, A. A. Clerk, Nature Communications {\bf 5}, 5562 (2014).

\bibitem{Jin2014}
J. Jin, M. Marthaler, and G. Sch\"on, Phys. Rev. B {\bf 91}, 085421 (2015).


\bibitem{Cottet2015}
A. Cottet, T. Kontos, and B. Doucot, Phys. Rev. B {\bf 91}, 205417 (2015).

\bibitem{Mendes2015}
U. C. Mendes and C. Mora, New J. Phys. {\bf 17}, 113014 (2015).

\bibitem{Quassemi2015}
F. Qassemi, A. L. Grimsmo, B. Reulet, and A. Blais, arXiv:1507.00322.

\bibitem{Hassler2015}
F.~Hassler and D.~Otten, Phys. Rev. B {\bf 92}, 195417 (2015).

\bibitem{Bajjani2010}
E. Zakka-Bajjani, J. Dufouleur, N. Coulombel, P. Roche, D. C. Glattli, and F. Portier,
Phys. Rev. Lett. {\bf 104}, 206802 (2010).

\bibitem{Pashkin2011}
Yu. A. Pashkin, H. Im, J. Lepp\"akangas, T. F. Li, O. Astafiev, A. A. Abdumalikov Jr., E. Thuneberg, and J. S. Tsai,
Phys. Rev. B {\bf 83}, 020502(R) (2011).

\bibitem{Reulet1}
G. Gasse, C. Lupien, and B. Reulet, Phys. Rev. Lett. {\bf 111}, 136601 (2013).


\bibitem{Reulet2}
J.-C. Forgues, C. Lupien, and B. Reulet,  Phys. Rev. Lett. {\bf 113}, 043602 (2014).


\bibitem{Reulet3}
J.-C. Forgues, C. Lupien, and B. Reulet,  Phys. Rev. Lett. {\bf 114}, 130403 (2015).


\bibitem{Portier2014} 
C. Altimiras, O. Parlavecchio, P. Joyez, D. Vion, P. Roche, D. Esteve, and F. Portier, Phys. Rev. Lett. {\bf 112}, 236803 (2014).


\bibitem{Delbecq2013}
M. R. Delbecq	L. E. Bruhat	J. J. Viennot	S. Datta,	A. Cottet, and T. Kontos, Nature Communications {\bf 4} 1400 (2013).


\bibitem{Rimberg2014}
F. Chen, J. Li, A. D. Armour, E. Brahimi, J. Stettenheim, A. J. Sirois, R. W. Simmonds, M. P. Blencowe, and A. J. Rimberg
Phys. Rev. B {\bf 90}, 020506(R) (2014).


\bibitem{Petta2015}
Y.-Y. Liu, J. Stehlik, C. Eichler, M. J. Gullans, J. M. Taylor, and J. R. Petta, Science {\bf 347}, 285 (2015).

\bibitem{Koch1} R.~H.~Koch, D.~J.~Van Harlingen, and J.~Clarke, Phys. Rev. Lett. {\bf 45}, 2132 (1980).


\bibitem{Koch2} R.~H.~Koch, D.~J.~Van Harlingen, and J.~Clarke, Phys. Rev. Lett. {\bf 47}, 1216 (1981).


\bibitem{Likharev}
K. K. Likharev, {\em Dynamics of Josephson Junctions and Circuits} (Gordon and Breach, New York, 1986).

\bibitem{Keldysh}
M.~Fleischhauer, J.~Phys.~A~{\bf 31}, 453 (1998).




\bibitem{Ingold1998}
H.~Grabert, G.-L.~Ingold, and B.~Paul, Europhys.~Lett.~{\bf 44}, 360 (1998).

\bibitem{Ingold1999}
G.-L.~Ingold and H.~Grabert, Phys.~Rev.~Lett.~{\bf 83}, 3721 (1999).

\bibitem{Wallquist2006}
M. Wallquist, V. S. Shumeiko, and G. Wendin, Phys. Rev. B {\bf 74}, 224506 (2006).

\bibitem{Leppakangas20142}
We note that in Refs.~[\onlinecite{Leppakangas2013,Leppakangas2014}]
we have made a sign error in the inductive term in the boundary condition of Eq.~(\ref{eq:BoundaryCondition}),
which means that all the equations there should be considered with changes $I_{\rm c}\rightarrow - I_{\rm c}$
and  $C_{\rm J}\rightarrow - C_{\rm J}$. This has no effect on any final result of these articles.

\bibitem{WallsMilburn}
D. F. Walls and G. J. Milburn, {\em Quantum Optics} (Springer, Berlin, 2008).
















\bibitem{QuantumNoiseBook}
C. Gardiner and P. Zoller, {\em Quantum Noise} (Springer, Berlin, 2004).


\bibitem{Loudon}
R. Loudon, {\em The Quantum Theory of Light} (Oxford University, New York, 2010).


\bibitem{SchoellerSchoen}
H. Schoeller  and G. Sch\"on,
Phys. Rev. B {\bf 50}, 18436 (1994).

\bibitem{Michael2015}
M.~Marthaler and J.~Lepp\"akangas, arXiv:1510.01985.

\bibitem{ThirdCumulant}
B. Reulet, J. Gabelli, L. Spietz, and D. E. Prober, arXiv:1001.3034.

\bibitem{Reulet2003}
B. Reulet, J. Senzier, and D. E. Prober, Phys. Rev. Lett. {\bf 91}, 196601 (2003).


\bibitem{GermanGuys}
C. Emary, C. P\"oltl, A. Carmele, J. Kabuss, A. Knorr, and
T. Brandes, Phys. Rev. B {\bf 85}, 165417 (2012).


\bibitem{Ingold2002}
H.~Grabert and G.-L.~Ingold, Europhys.~Lett.~{\bf 58}, 429 (2002).



\bibitem{Pozar}
D. M. Pozar, {\em Microwave Engineering, 2nd ed.} (Wiley, New York, 1998).



\end{thebibliography}
\end{document}